\documentclass[onecolumn,useAMS]{mn2e}
\usepackage{epsfig}
\def\la{\lower.5ex\hbox{$\; \buildrel < \over \sim \;$}}
\def\ga{\lower.5ex\hbox{$\; \buildrel > \over \sim \;$}}
\def\apj{ApJ}
\def\mnras{MNRAS}
\def\prd{PRD}
\def\apjl{ApJL}
\def\nat{Nature}
\def\aj{AJ}
\def\aap{A \& A}
\def\araa{Ann. Rev. of  Astronomy \& Astrophysics}
\def\physrep{Physics Reports}
\begin{document}

\title[Reionization by Primordial Magnetic fields]{Primordial Magnetic
Fields in the Post-recombination Era and Early Reionization }
\author[Shiv K. Sethi and Kandaswamy Subramanian]
{Shiv K. Sethi$^1$ and  Kandaswamy Subramanian$^2$  \\
\hspace{-0.1cm} ${}^1$Raman Research Institute, Bangalore 560080, India \\
\hspace{-0.1cm} ${}^2$Inter-University Centre for Astronomy and Astrophysics,
 Pune, India \\
\hspace{-0.1cm} emails: sethi@rri.res.in, kandu@iucaa.ernet.in
}

\maketitle
\begin{abstract}
We explore the ways in which primordial magnetic fields influence 
the thermal and ionization history of the post-recombination universe.
After recombination the universe
becomes mostly neutral resulting also in a sharp drop in the 
radiative viscosity.  Primordial magnetic fields can then dissipate
their energy into the intergalactic medium (IGM) via ambipolar diffusion and, 
for small enough scales, by generating decaying MHD turbulence.
These processes can significantly modify the thermal and ionization history of 
the post-recombination universe. We show that the dissipation
effects of magnetic fields which 
redshifts to a present value $B_{0}=3\times 10^{-9}$ Gauss
smoothed on the magnetic Jeans scale and below, can
give rise to Thomson scattering optical depths $\tau \ga 0.1$,
although not in the range of redshifts needed to explain the
recent WMAP polarization observations. We also study  the 
possibility that primordial fields could  induce the formation 
of subgalactic structures for $z \ga 15$.  
We show that early structure formation induced by nano-Gauss magnetic fields
is  potentially capable of producing the early
re-ionization implied by the WMAP data. Future CMB
observations will be very useful to probe the modified
ionization histories produced by primordial
magnetic field evolution and constrain their strength.

\end{abstract}

\section{Introduction}

Magnetic fields play an important role in understanding most structures 
in the universe (Parker 1979; Zeldovich, Ruzmaikin \& Sokoloff 1983). 
Their origin however is not yet clearly understood. Observed galactic 
magnetic fields could have arisen from dynamo amplification 
of a seed magnetic fields $\simeq 10^{-20} \, \rm G$
(cf. Ruzmaikin, Shukurov \& Sokoloff 1988; Beck et al 1996; Shukurov 2004;
Brandenburg \& Subramanian 2004); with the seed field itself 
originating in the early
universe or from astrophysical processes (e.g. Harrison 1970; 
Subramanian, Narasimha and Chitre 1995; Kulsrud et al 1997; Grasso \&
Rubenstein 2001 and Widrow 2002 for reviews).
However there are potential difficulties
for the dynamo theory to overcome, due to the constraints implied by
helicity conservation and the more rapid growth of small-scale magnetic
fields (Cattaneo \& Vainshtein 1991; Kulsrud \& Anderson 1992;
Gruzinov \& Diamond 1994; Subramanian 1998, 1999, 2002;
Blackman \& Field 2000; Kleoorin {\it et al} 2000;
Brandenburg 2001; Brandenburg \& Subramanian 2000;
Blackman \& Brandenburg 2002;
and Brandenburg \& Subramanian 2004 for a recent review).
Magnetic fields with larger coherence scales may also be present in
clusters of galaxies (Clarke, Kronberg \& Bohringer 2001; Carilli and
Taylor 2002; Vogt and Ensslin 2003) and at high redshifts 
(Oren \& Wolfe 1995).  Such large-scale coherent fields 
potentially present further problems for the dynamo paradigm.

Alternatively, large-scale magnetic fields could also have
arisen from primordial magnetic fields $\simeq 10^{-9} \, \rm G$, 
generated in the early universe, for instance during inflation (cf. 
Turner \& Widrow 1988; Ratra 1992; cf. Grasso \& Rubenstein 2001;
Giovannini 2004 for reviews). Such a primordial tangled magnetic field 
will also influence the formation of structures in the universe, like
galaxies (Rees \& Reinhardt 1972; Wasserman 1978; Kim,
Olinto \& Rosner 1996; Subramanian \& Barrow 1998a, SB98a hereafter, 
Sethi 2003) and give rise to CMBR temperature and 
polarization anisotropies (Barrow, Ferreira \& Silk
1997; Subramanian \& Barrow 1998b, 2002; Durrer, Ferreira, 
Kahniashvili, 2000; Seshadri \& Subramanian 2001; 
Mack, Kashniashvili, Kosowsky, 2002; Subramanian, Seshadri \& Barrow 2003). 
These considerations can be used to constrain the magnetic field
amplitude and the shape of its power spectrum. In this paper we 
consider the possible role magnetic fields could have played
in determining the thermal and ionization history of the universe in 
the post-recombination epoch. 

For redshifts $z \la 1100$, primeval plasma begins to recombine to form neutral
hydrogen. The ionized fraction  decreases by nearly an order of 
magnitude by $z \simeq 1000$, finally reaching a value of $\simeq 10^{-4}$
for $z \la 100$ (for details see Peebles 1993). The matter temperature
continues to follow the CMBR temperature, both falling as $\propto 1/a$ for
$z \ga 100$. (here $a$ is the expansion factor of the universe).
At smaller redshifts, matter 'thermally' decouples from the 
radiation and the matter temperature falls as $\propto 1/a^2$. In the 
standard picture, this  thermal and ionization history holds up to $z \simeq
10\hbox{--}20$, when the formation of first structure can lead to 
reionization and reheating. Recent WMAP observations of CMBR anisotropies
suggests that the universe reionized as early as $z \simeq 17$ (Kogut et al.
2003). On the other hand, observations of high redshift quasars suggest
that the universe is fully ionized for $z \la 5$ but the neutral
fraction reaches values $\simeq 10\%$ between $5.2 \la z \la 6$ (Fan et al.
2002, Becker et al. 2001, Djorgovski et al. 2001).   
Such early reionization, as implied by WMAP observations,
presents a challenge, and is not yet fully understood (cf. Ricotti \&
Ostriker 2003). One of our aims will also be to explore
the possibilities offerred by primordial magnetic fields
in this respect.

In the pre-recombination epoch, magnetic fields evolve as 
$B \propto 1/a^2$ on sufficiently large scales, larger than
the magnetic Jeans length $\lambda_J$ (see below). 
In the post-recombination epoch
there is a sharp drop in the electron density as the 
universe becomes mostly neutral. This can lead to dissipation of 
magnetic field energy into the medium from ambipolar diffusion 
(Cowling 1956, for details see Shu 1992 and references therein). 
There is also a sharp drop in radiative viscosity after 
recombination (Jedamzik, Katalini{\' c}, \& Olinto 1998; SB98a).  
For scales smaller than $\lambda_J$, non-linear effects
can then lead to the field generating decaying MHD turbulence,
and the dissipation of magnetic field energy on such scales, 
into the intergalactic medium. These processes will affect the
thermal and ionization history of the universe.
In this paper we explore how the standard ionization and thermal 
history might get modified due to such dissipation of magnetic field 
energy in the post-recombination epoch, from ambipolar diffusion and decaying 
turbulence. 

In addition tangled primordial magnetic fields
can also induce early formation of structures in the universe 
(Kim et al. 1996; SB98, Gopal \& Sethi 2003).
We study in detail this process 
and find the range of redshifts at which the first structures might collapse. 
Such early structure formation can also lead to changes in 
the thermal and ionization history of the universe, and be
a potential source of the early re-ionization inferred from WMAP.

In the next section we set the notation for describing
primordial magnetic fields and summarize briefly
some of the relevant results from earlier work on their evolution.
The processes which dissipate
magnetic field energy into the IGM are considered in \S 3.
After briefly describing the equations that govern the ionization and 
thermal history of the universe in \S 4, we apply them to
compute  the effect of magnetic energy dissipation on the IGM
in the subsequent section. The formation of the first nonlinear
structures induced by tangled primordial magnetic fields
and their possible effects are considered in \S 6,
The last section summarizes our conclusions.  Throughout this paper we 
use, unless specified otherwise, 
 the currently-favoured FRW  model: spatially flat
with $\Omega_m = 0.3$ and $\Omega_\Lambda = 0.7$ (Spergel et al. 2003, Reiss et al. 2004, Tonry et al. 2003, Perlmutter  et al. 1999,
Riess  et al. 1998) with  $\Omega_b h^2 = 0.02$ (Spergel et al. 2003, Tytler 
 et al. 2000) and 
$h = 0.7$ (Freedman  et al. 2001). 

\section{Magnetic Fields in the early universe}

Let us suppose that some process in the early
universe generated tangled magnetic fields.
Firstly, on large-enough scales, the velocity induced by tangled
primordial fields is generally so small that it does not lead to any
appreciable distortion of the initial field (Jedamzik, Katalinic and Olinto
1998, SB98a). In this 'linear' regime, to a very
good approximation, the magnetic field simply redshifts  as ${\bf B}(%
{\bf x},t)=\tilde{\bf B}({\bf x})/a^{2}$. Here ${\bf x}$ is the comoving
coordinate. We assume the early universe tangled magnetic field,
$\tilde{\bf B}$ to be initially an isotropic and homogeneous random process.
This allows one to write, 
in Fourier space (see .i.e. Landau \& Lifshitz 1959):
\begin{equation}
\langle {\tilde{B}_i({\bf q})\tilde{B}^*_j({\bf k})} \rangle 
= \delta^3_{\scriptscriptstyle D}({\bf q -k}) 
\left (\delta_{ij} - k_i k_j/k^2 \right ) M(k)
\label{eq:n6}
\end{equation}
Here $M(k)$ is the magnetic field power spectrum and 
$k=\vert {\bf k}\vert$ is the comoving wavenumber. 
This gives $\langle \tilde{\bf B}^2({\bf x}) \rangle 
=\int (dk/k) \Delta_B^{2}(k)$, where $%
\Delta_B^{2}(k)= 8 \pi k^{3}M(k)$ is the power per logarithmic
interval in $k$ space residing in magnetic tangles,

We shall often consider the following two forms
of the power spectrum $M(k)$. For several applications,
the results are dominated by the smallest scales of the
tangled magnetic field. It then suffices to consider
a simple, single scale power spectrum:
\begin{equation}
M(k) =  B_0^2 \delta(k -k_\star) /(8\pi k^2)
\label{eq:mag_pow}
\end{equation}
In this  normalization  $B_0$ coincides with the RMS of the magnetic
field, redshifted to the present epoch i.e.
$ B_0^2 = \langle \tilde{\bf B}^2({\bf x}) \rangle$.
We shall also consider the effects of taking a more
complicated spectrum, like a power law spectrum,
with $M(k) = Ak^n$ cut off at $k=k_{max}$; with $k_{max}$
determined by the effects of damping by radiative viscosity
before recombination. One can fix $A$ by demanding
that the smoothed field strength over a
scale, $k_{G}$, (using a sharp $k$-space
filter) is $B_{G}$. We will assume $n>-3$. Using the same filter, the RMS
value of the field smoothed over a wavenumber $k$, is given by 
\begin{equation}
\tilde{B}^2(k)= B_G^2 (k/k_{G})^{3+n}.  
\label{eq:powspec}
\end{equation}

As we shall see below
the energy dissipation from ambipolar diffusion depends on 
$\langle ({\bf \nabla}_{\bf r} \times {\bf B}) \times  {\bf B})^2 \rangle
= (1/a^{10}) \langle ({\bf \nabla}_{\bf x} \times \tilde{\bf B}) \times  
\tilde{\bf B})^2 \rangle$, where now ${\bf r} = a{\bf x}$ is the proper 
co-ordinate. Using Eq.~(\ref{eq:n6}) and assuming the magnetic fields 
to be a Gaussian process:
\begin{equation}
\langle ({\bf \nabla}_{\bf x} \times \tilde{\bf B}) \times  
\tilde{\bf B})^2 \rangle = {7 \over 3} \int dk_1 
\int dk_2 M(k_1)  M(k_2) k_1^2 k_2^4
\label{eq:n6pp}
\end{equation}

The effects of the distortion of the initial
magnetic field, and the nonlinear processing of the
magnetic spectrum begins to be felt at all
scales $l$ which satisfy the inequality $v(l)/l \ga H(t)$.
Here $v(l)$ is the velocity induced by the magnetic field
on a proper scale, $l = a/k$, and $H=\dot{a}/a$ is the Hubble expansion
rate at cosmic time $t$ (cf. Banerjee and Jedamzik 2003; SB98a). 
For large fluid Reynolds number (see below),
the fluid velocity induced by the tangled magnetic field is
of order the Alfv\'en velocity; that is $v(l) \approx V_A(k=a/l,t)$
where in the post-recombination epochs, $V_A(k,t)$ is given by
\begin{equation}
V_A(k,t) = {\frac{B(k,t)}{(4\pi \rho _{b}(t))^{1/2}}}
\approx 1.5\times 10^{-5}\left({ B(k,t) a^2 \over 10^{-9}{\rm G}}\right) 
\  c \ a^{-1/2}.
\label{alfven}
\end{equation}
Here $\rho_b(t)$ is the density of baryons,
and $B(k,t)$ is the magnetic field smoothed on
a scale $l=a/k$ at time $t$. For scales which are
not yet affected by nonlinear processing, one can assume the field
evolves approximately with a constant $B(k,t) a^2(t) = \tilde{B}(k) = 
B_G (k/k_G)^{(n+3)/2}$, where for the second equality
we have adopted the power law spectrum of Eq.~\ref{eq:powspec}. 

Note that the scale below which nonlinear effects become important
is also approximately equal to the
magnetic Jeans length, below which
the distortion of the field can lead to
magnetic pressure gradients which can counteract the gravitational
collapse (cf. SB98a).
In fact in a linear analysis this condition allows us to define 
the proper magnetic Jeans' wave number, say $K_J$, from 
equating the two terms: $4 \pi G \rho_m =  
K_J^2 B^2/(8 \pi\rho_b)$, giving 
\begin{equation}
K_J = {4\pi \sqrt{2\rho_m \rho_b G} \over B}. 
\end{equation}
Defining the comoving Jeans scale
by $k_J = aK_J$, and noting that $H^2(t) = 8\pi G\rho_m/3$,
the above condition is equivalent to
the condition $[k_J/a(t)] V_A(k_J,t) = \sqrt{3} H(t)$.
This can be explicitly seen to be very similar to 
the condition discussed above on the length scales
below which nonlinear evolution is important.
The comoving Jeans' length 
$ \lambda_J= 2\pi/k_J$,  or the comoving Jeans' wavenumber 
$k_J$ do not depend on time,
at early epochs where the universe is matter dominated; and 
assuming that the field even at the scale $k_J$
just redshifts as $\propto 1/a^2$,
without significant distortion (cf. SB98a). This is because
in this case $V_A \propto a^{-1/2}$ and $H(t) \propto t^{-1} \propto a^{-3/2}$
and hence $k_J \propto a(t)H(t)/V_A$ is constant with time.
So any scale which is linear/nonlinear just after recombination,
is approximately linear/nonlinear at all epochs 
(till the vacuum energy starts dominating).

Putting in numerical values we get
\begin{equation}
k_J \simeq 14.8\, {\rm Mpc^{-1}} 
\left ({\Omega_m \over 0.3 } \right )^{1/2} \left ({h \over 0.7} 
\right )  \left ({B_J \over 10^{-9} {\rm G}} \right )^{-1}
\end{equation}
where $B_J = B(k_J,t) a^2(t)$ is the redshifted
value of the field smoothed on the scale $k_J$.
Again for the power law spectrum given by 
Eq.~\ref{eq:powspec}, we will have $B_J = B_G(k_J/k_G)^{(n+3)/2}$,
giving an implicit equation for $k_J$. For a nearly scale
invariant spectrum with say $n=-2.9$, $k_G = 1h \, \rm Mpc^{-1}$, 
one gets for the above cosmological parameters
$k_J \sim 13 \, {\rm Mpc^{-1}} (B_G/10^{-9} {\rm G})^{-0.95}$. 

There is another scale which plays an important role
in what follows. This is the Alfv\'{e}n-wave damping length-scale 
$k_{\rm max}$, below which tangled magnetic fields are strongly damped
by radiative viscosity in the pre-recombination universe 
(Jedamzik, Katalinic and Olinto 1998, SB98a). We have 
\begin{equation}
k_{\rm max} \simeq  235 \, {\rm Mpc^{-1}} \left ({B_m 
\over 10^{-9} {\rm G}} \right )^{-1} 
\left (\Omega_m \over 0.3 \right )^{1/4} 
\left ( \Omega_b h^2 \over 0.02 \right )^{1/2} 
\left ( h \over 0.7 \right )^{1/4}
\label{eq:kmaxn}
\end{equation}
Here $B_m$ is the field smoothed over scales 
larger than the cut-off scale,
(and redshifted to the present epoch), which act
as the effective large scale field for the cut-off scale 
perturbations (cf. Jedamzik, Katalinic \& Olinto 1998; SB98a;
Seshadri \& Subramanian 2001).

The magnetic field smoothed 
on comoving wave numbers $k \la k_J$ evolve in a 'linear'
fashion in the post-recombination epoch. As we discuss below
the magnetic field at these scale is 
mainly dissipated by ambipolar diffusion. 
For $k_{\rm max} \ga k \ga  k_J$, 
non-linear effects can also lead to decaying MHD turbulence and
consequent dissipation of the magnetic field energy.
In the post-recombination era density perturbations seeded by the 
primordial magnetic field can grow on scales 
with $k \la k_J$ (Wasserman 1978; Kim et al 1996; SB98a; 
Sethi 2003; Gopal \& Sethi 2003). 
This can lead to early collapse of structure which might have important 
implication for the ionization of the universe. We discuss this scenario
in more detail in \S 6. Before this we first consider the
magnetic field energy dissipation into the IGM due 
to ambipolar diffusion and decaying turbulence.

\section{Energy input into the IGM from primordial magnetic field dissipation}

\subsection{Ambipolar diffusion}

Ambipolar diffusion is important for magnetic field energy decay 
in a mostly neutral medium. The post-recombination universe satisfies 
this criterion as the ionized fraction of hydrogen 
$n_e/n_{\rm \scriptscriptstyle B} \equiv 
 x_e \simeq 10^{-4}$ at $z \simeq 100$
(Peebles 1968, Zel'dovich, Kurt, \& Sunyaev 1969, Peebles 1993). 
In the presence of a tangled magnetic field, the 
Lorentz force acts only on the small fraction of ionized component, thereby 
generating a velocity difference between the ionized and the neutral components.
This relative  velocity between charged and neutral particles 
is damped by ion-neutral collisions, which
leads to  dissipation of magnetic field energy (see e.g. Cowling 
1956, Mestel \& Spitzer 1956). This energy dissipation
process known as ambipolar diffusion is important 
in mostly neutral molecular clouds (for details see Shu 1992).

The  neutral particles  in the early universe are neutral hydrogen 
and helium. Throughout this
paper we neglect the effect of energy dissipation on the small 
fraction of  neutral helium atoms. The volume rate of energy dissipation 
due to ambipolar diffusion is then (Cowling 1956)
\footnote{Note that the expression below doesn't agree with 
Shu (1992) Eq.~(27.19), though the 
formulae agree for a  mostly neutral medium. The formula given
here is  correct for an arbitrary ionized medium (Cowling (1956), Eq.~(27))}:
\begin{equation}
\Gamma_{\rm in} = {\rho_n \over 16 \pi^2 \gamma \rho_b^2 \rho_i} 
|({\bf \nabla x B) x B}|^2
\label{eq:amdif}
\end{equation} 
Here $\rho_n$, $\rho_i$, and $\rho_b$ are the densities of 
neutral hydrogen, ionized hydrogen, and total baryon density,  respectively. 
Also $\gamma = \langle w \sigma_{\rm in} \rangle /(m_n + m_i)$ (Shu 1992); 
where $w$ is the ion-neutral relative velocity and $\sigma_{\rm in}$ is the 
cross section for the collision between ions and neutrals. 
For $w \la 10 \, \rm km \, sec^{-1}$, 
$\langle w \sigma_{\rm in} \rangle \simeq 3 \times 10^{-9}$ independent 
of the relative velocity of ions and neutrals. This
approximation  holds for the parameter space we consider in this paper 
(for a detailed discussion and references 
see Shu 1992). This energy is deposited into the neutral
component of the medium. However owing to collisions between electrons,
protons, and neutrals the energy is rapidly thermalized at rates 
much higher than  the expansion rate of the universe (see e.g. Madau, Meiksin
\& Rees 1997 and references therein). The volume 
rate of energy deposition in electrons, 
(required for Eq.~(\ref{eq:elect}) below),  
is $\Gamma_e = x_e \Gamma_{\rm in}$.
Ambipolar diffusion is the main process of magnetic field energy dissipation
at comoving length scales at which velocities are linear .i.e. 
$k \la k_J$.
However, even smaller length scales, upto $k^{-1} \la k^{-1}_{\rm max}$ 
can contribute to the energy input into the IGM due to
ambipolar diffusion (if the decay due to the the MHD
turbulence discussed below is not efficient enough).

\subsection{Decaying Turbulence}
\label{decay}

For magnetic fields which vary on length scales smaller
than the magnetic Jeans scale, or for $k > k_J$, an additional mode
of decay is possible. Such small scale tangled magnetic fields can induce
decaying MHD turbulence. 

Firstly, as we pointed out in \S 2,
for magnetic fields at comoving scales below the comoving
magnetic Jeans' length i.e. $k > k_J$,
the rate of energy transfer due to the nonlinear interaction between modes, 
$\sim (k/a)V_A(k,t)$ becomes smaller than the Hubble rate $H(t)$.
However, just prior to recombination,
velocity perturbations at these scales,
are over-damped  owing to large radiative viscosity of the medium
(Jedamzik, Katalini{\' c}, \& Olinto 1998, SB98a)
and therefore cannot lead to turbulence. 
After recombination, the radiative viscosity dramatically
decreases; the viscous force per unit mass 
${\bf F}_V/\rho_b$ becomes that due to the free
streaming photons, with ${\bf F}_V/\rho_b = - \alpha {\bf v}$, where
$\alpha = (4/3)(\rho_{\gamma}/\rho_b) n_e\sigma_t c$.
Here $\sigma_t$ is the Thomson cross section and $\rho_{\gamma}$
the photon density. The corresponding Reynolds number for fluid motions
(The dimensionless ratio of the nonlinear term 
${\bf v}\cdot{\bf \nabla}{\bf v}$ to the viscous term in the
fluid momentum equation), which is given by 
$R = kv/(a\alpha)$ becomes very large. In fact
writing $R= (k/k_J)(v/V_A) R_J$,
we have
\begin{equation}
R_J = {3\sqrt{3} \rho_b \over 4 \rho_\gamma} {L_\gamma \over D_H}
\label{reynolds}
\end{equation}
where $L_\gamma= (n_e\sigma_t a)^{-1}$ is the comoving photon mean 
free path and $D_H = c/(Ha)$ is the comoving Hubble radius. 
We have used also the fact that
$[k_J/a(t)] V_A(k_J,t) = \sqrt{3} H(t)$ in deriving Eq.~\ref{reynolds}.
Just after recombination,
$L_\gamma \sim 10^4$ Mpc (for an ionization fraction of $10^{-4}$)
while $D_H \sim 100$ Mpc. So magnetic tangles with 
$k > k_J$, for which nonlinear 
interactions are important, the Reynolds numbers are also
large enough that decaying MHD turbulence can be induced. 
This becomes increasingly the case at smaller redshifts, because
$R_J \propto a^{5/2}$ increases rapidly as the universe expands.

It is also important to check whether the ambipolar diffusion
discussed above allows for such turbulence.
For this one needs to estimate the magnetic Reynolds number
associated with ambipolar diffusion, that is the dimensionless ratio 
of the advection term in the induction equation for ${\bf B}$,
to the ambipolar diffusion term. This is given by
$R_{\rm ambi} \sim K v B/ [K^2 B^2/(4 \pi \rho_i \rho_n \gamma)]$,
where $K = k/a$ is the proper wavenumber and $v$                  
the velocity induced by the magnetic field. We have numerically,
\begin{equation}
R_{\rm ambi} \approx  
8.5 \times 10^3 \frac{v}{V_A} \left(\frac{1+z}{10^3}\right)^{3/2} 
\left(\frac{x_e}{10^{-3}}\right) \left(\frac{k}{k_J}\right)^{-1}
\label{rambi}
\end{equation}
So initially we have $R < R_{\rm ambi}$ and radiative viscosity is
more important; we have already seen that this itself is
weak enough to allow decaying turbulence to be induced.
As the universe expands and with decreasing redshift, $R_J$ increases as 
$a^{5/2}$, while $R_{\rm ambi}$
decreases as $a^{-3/2} \propto (1+z)^{3/2}$ and so eventually ambipolar
diffusion is more important. But even for $k=k_{max}=235$, 
$R_{\rm ambi}$ is of order unity only at redshifts $z \sim 20$.
So for modes with $ k > k_J$
decaying turbulence is important from recombination to
fairly low $z < 20$ (for $k < k_{max}$), after which
any remaining energy will be drained much faster 
by ambipolar diffusion.  For decaying turbulence,
nonlinear interactions between different modes
causes the magnetic field to decay by cascading energies to 
smaller and smaller scales and subsequent dissipation, 
independent of the exact physical mechanism of dissipation   
\footnote{Note that for smaller scales, $R$ due to the free-streaming
radiative viscosity will in fact increase. However for smaller
scales the diffusive fluid viscous force and ambipolar damping
becomes increasingly more important. The corresponding Reynolds
number due to fluid viscosity is $R_f = av/k\nu$, where
$\nu$ is the kinematic viscosity of the hydrogen gas.
And that due to ambipolar damping is $R_{\rm ambi}$ given in the text.
Just after recombination, both $R_f(k_J) \gg R_{\rm ambi} \gg R(k_J)$.
But since $R_f$ and $R_{\rm ambi}$ decrease with increasing $k$,
while $R$ increases with $k$, for a large enough $k$
we will have $R_{\rm ambi}(k) < R(k)$. Below this scale
ambipolar damping will dominate the radiative viscosity, and drain
the energy from the turbulence.
The energy lost from the magnetic field thus eventually
goes to heat the gas.}.

Simulations of such decaying MHD turbulence in flat space suggests that
the magnetic energy decays as a power law at large times 
(see e.g. Banerjee \& Jedamzik 2003, 
Christensson, Hindmarsh \& Brandenburg 2001, 
Biskamp \& M{\"u}ller 2000), the exact power law
depending on the magnetic spectral index on large scales.
One can model this decay as, 
\begin{equation}
{\cal E}_{\rm B} = {{\cal E}_{\rm B0} \over (1 + \tilde t/\tilde t_{d})^m}
\label{eq:flatdec}
\end{equation}
where $\tilde t$ is the time in 'flat-space',
$\tilde t_{d}$ is the relevant dynamical time for decay,
and ${\cal E}_{\rm B}$ is the magnetic energy in flat space,
with ${\cal E}_{\rm B0}$ its initial value. Also simple
scaling arguments suggest $m = 2(n+3)/(n+5)$, for
an initial power spectrum with $n > -3$ (cf. Olesen 1997, Shiromizu 1998,
Christensson, Hindmarsh \& Brandenburg 2001, 
Banerjee \& Jedamzik 2003).
One can take over this result into the
expanding universe if one can map the flat space 
MHD turbulence decay problem into
the FRW universe. It is well known (Brandenburg, Enqvist \&  Olsen 1996, 
SB98a) that viscous MHD in the
radiation dominated universe can be conformally mapped to
flat space viscous MHD. In the matter dominated era the following 
transformation of variables can be used to approximately map the 
expanding universe viscous MHD to flat space (Banerjee 2003):
\begin{equation}
\tilde{\bf B} = {\bf B} a^2, \quad 
d\tilde t = dt/a^{3/2}, \quad \tilde{v} = a^{1/2} v, \quad 
\tilde \rho = \rho a^3, \quad \tilde p = p a^4, \quad 
\tilde \epsilon = a^4 \epsilon
\end{equation}
The validity of this mapping  requires $p \ll \rho$.
Integrating $d\tilde t = dt/a^{3/2}(t)$, with $a^{-3/2}(t) = c_1/t$,
as appropriate for a matter dominated universe (which obtains
till the vacuum energy dominates),
one gets $\tilde t = c_1\ln(t/t_i)$; here we have also
used the condition that
$\tilde t = 0$ when $t=t_i$ the initial epoch when the decay starts,
Similarly $\tilde t_d = c_1 \ln((t_i + t_d)/t_i)$, where $t_d$
is the physical decay timescale for the turbulence, which we
may approximate as the Alfv\'en time scale for the mode $k_{max}$.
So $\tilde t/\tilde t_d = \ln(t/t_i)/\ln(1+(t_d/t_i))$ independent
of the constant $c_1$. 
Further, using the above transformation for the magnetic field,
$\tilde{\bf B} = {\bf B} a^2$, the magnetic energy will scale as
$E_B(t) = {\cal E}_B/a^4$, where ${\cal E}_B$ itself, in our
phenomenological model, decays as in Eq.~(\ref{eq:flatdec}).
The transformation between $\tilde t$ and $t$
then implies that a power-law decrease
in magnetic field energy in flat space 
will correspond to a logarithmic decrease of the energy
(on scales $ k_J < k < k_{max}$), in the 
matter dominated era, due to decaying
turbulence, over and above the decay due to expansion. 
Also using this transformation 
we can model the rate of dissipation of magnetic field
energy, on these scales, in a matter dominated  expanding universe as:
\begin{equation}
{dE_{\rm B} \over dt} = -4 H(t) E_B - {3m H(t) E_B \over 
2 [ \ln(1 + t_d/t_i) + \ln(t/t_i) ]}
\label{decaymod}
\end{equation}
Here the first term corresponds the redshifting due to expansion
and the second due the energy lost from the field due to the
decaying turbulence. 
Using   $E_B(t) = {\cal E}_B/a^4$ and substituting
for ${\cal E}_B$ from Eq.~\ref{eq:flatdec}, we  have for the rate of
energy input due to decaying turbulence,
\begin{equation}
\Gamma_{\rm decay} = {B_0(t)^2 \over 8\pi}{3m \over 2}
{[\ln( 1 +t_d/t_i)]^m \over [ \ln(1 + t_d/t_i) + \ln(t/t_i) ]^{m+1}}
\times H(t),
\label{eq:dectu}
\end{equation}
where we have defined $B_0^2(t)/8\pi = {\cal E}_{B0}/a^4$.
For $(t-t_i) \ll t_d$, the second term in the denominator, 
$\ln(t/t_i)$, can be dropped. 
On the other hand, for $t-t_i \gg t_d$, the first term, $\ln(1 + t_d/t_i)$ in
the denominator of Eq.~\ref{eq:dectu} can be dropped, to get
\begin{equation}
\Gamma_{\rm decay} = {B_0(t)^2 \over 8\pi}{3m \over 2}
{[\ln( 1 +t_d/t_i)]^m \over 
\ln^{m+1}\left[((1+z_i)/(1+z))^{3/2} \right ]} \times H(t)
\label{eq:dectuL}
\end{equation}
Here we have used the relation $t \propto a^{3/2} \propto (1 +z)^{-3/2}$
valid in the matter dominated universe at high redshift.
We are now in a position to study the effects of the energy
input into the IGM due to ambipolar diffusion and decaying turbulence;
first we recall briefly the relevant equations involved
in calculating the thermal and ionization history of the IGM.

\section{Thermal and Ionization state of IGM}
In the post-recombination universe, the IGM is mostly neutral reaching 
an ionization fraction $x_e \simeq 10^{-4}$ by $z \simeq 100$ (Peebles
1993). At smaller 
redshifts the ionization fraction decreases very slowly because the recombination rate of hydrogen  is much smaller than the expansion rate of the 
universe with such small ionization levels. The temperature of matter (which
refers to the temperature of either electrons, protons, or the neutral particles which remain in equilibrium with each other) continues to be nearly  equal to the  CMBR temperature for $z \ga 100$ because of inverse Compton 
scattering (see e.g. Peebles 1993). At smaller redshifts the matter temperature $T_e \propto 1/a^2$, i.e. it falls more rapidly as compared to CMBR temperature $T_{\rm cbr}$  
which evolves  as  $\propto 1/a$. The dissipation of magnetic field energy 
into the IGM  alters its  thermal and ionization evolution. It  heats 
 the system which in turn  changes the ionization state by
collisional ionization. Thermal and ionization evolution in the presence of 
magnetic field dissipation is given by (see e.g. Peebles 1993):
\begin{eqnarray}
\dot T_e   & = & -2 {\dot a \over a} T_e + {x_e \over 1 + x_e} 
{8 \rho_\gamma \sigma_t \over 3 m_e c } (T_\gamma - T_e) 
+  { \Gamma_e  \over (1.5 k_B n_e)}  \nonumber  \\
\dot x_e & =  & \left [ \beta_e (1-x_e)\exp\left(-h\nu_\alpha/(k_BT_{\rm cbr})\right )-\alpha_e n_b x_e^2 \right ] C + \gamma_e n_b (1-x_e)x_e
\label{eq:elect}
\end{eqnarray}
Here $\Gamma_e$ is the volume rate of injection of 
energy into the electrons. $\Gamma_e = x_e \Gamma_{\rm in}$  for energy
dissipation from ambipolar diffusion (Eq.~(\ref{eq:amdif})) and 
$\Gamma_e = x_e \Gamma_{\rm decay}$ for energy injection from 
MHD decaying turbulence (Eq.~(\ref{eq:dectu})). 
(Also $k_B$ is the Boltzmann constant). In the ionization 
 evolution equation  the first two terms are the usual terms for the 
recombination of the primeval plasma (for details and notation
 see Peebles 1968, Peebles 1993). For 
$z \la 1000$, $C \simeq 1$ and the first term on the right hand of the equation
rapidly decreases. This means that after the recombination is completed the 
only important term is the recombination term (the second term on the right
hand side) which gives a slow decrease in the ionization fraction. The third
term on the right hand side corresponds to the collisional ionization 
of the IGM. This term is usually very small because collisional ionization
coefficient $\gamma_e$ is exponentially suppressed at  temperatures in the 
post-recombination universe 
 $T_e \simeq 0.25 ((1+z)/1000) \, \rm eV$, 
which are  much smaller than  the hydrogen 
ionization potential, $E_i = 13.6 \, {\rm eV}$.  Owing to magnetic
field energy dissipation, the matter temperature might  increase 
sufficiently  to make the collisional ionization term important and thereby
cause  a substantial change  in the ionization state of IGM. We show that this 
mechanism can be responsible for partially ionizing the universe at high
redshifts. 

\section{Thermal and ionization history changes due to primordial
magnetic field dissipation }

\subsection{The effects of ambipolar diffusion}

As already discussed, ambipolar diffusion is 
the principle mechanism of the dissipation of magnetic field energy
for magnetic field at scales $k \la k_J$. At smaller scales
decaying MHD turbulence can also dissipate energy, but ambipolar 
diffusion might  continue to be an important source 
of energy dissipation for $k \simeq k_{\rm max}$. 
(Indeed, we find that less than 40 percent of the energy is dissipated
by decaying turbulence for n=-2.9 index even by $z=10$;
so for nearly scale invariant spectra it would be appropriate also 
to consider ambipolar diffusion effects for scales up to $k_{max}$).

One can get a rough estimate  of the values of $B_0$ which might
cause significant change in ionization and thermal history of the universe  by   comparing the magnetic  field energy density with the 
thermal energy density of the universe. If at a certain epoch 
a fraction $f$ of magnetic field energy is dissipated into the 
IGM then it will typically raise it to a temperature: 
$T = f B_0^2/(8 \pi)/n_b k_B$; 
with $n_b = n_b(t_0) (1+z)^3$ and $B_0 = B_0(t_0)(1+z)^2$. Taking 
$f = 0.1$, this give $T \simeq 10^4 \, \rm K [(1+z)/100]$ for 
$B_0 \simeq 10^{-9} \, \rm G$. For $z \ga 100$, this
is an overestimate because owing to inverse
Compton scattering off CMBR photons matter temperature cannot increase 
much above CMBR temperature (Eqs.~(\ref{eq:elect})).  However 
it does give  a rough estimate of the magnitude of $B_0$ 
that are of 
interest. The fraction of the energy dissipated $f$  will depend on the 
magnetic field power spectrum.
 From Eq.~(\ref{eq:n6pp}) and Eq.~(\ref{eq:amdif}), 
it can be seen that the rate of dissipation is dominated by the smallest
scale (largest $k$) for the scale-free
magnetic field power spectrum (Eq.~(\ref{eq:powspec});  
This will correspond to the large-k cut-off, $k_{\rm max}$.

In Figure~\ref{fig:f1} and~\ref{fig:f2} we show ionization and 
thermal history of the universe
for some interesting values of $B_0$ for both delta function and power
law power spectra.  For delta function power spectrum, we take
$k_\star = k_{\rm max}$ (Eq.~{\ref{eq:kmaxn}) and for the power law 
power spectrum $B_0$ is 
defined as RMS value smoothed at $k= k_{\rm max}$. For integrating 
Eqs.~(\ref{eq:elect}) we start with initial conditions $T_e = T_{\rm cbr}$ and 
$x_e = 1$ at $z = 2000$. For ambipolar diffusion, we do not take into account 
the back-reaction of the energy dissipation on the magnetic field energy
and use $B \propto 1/a^2$.  We
are justified in neglecting it as  for values of $B_0$ that 
are   of interest less than 10\% of the magnetic field energy is dissipated 
into the IGM by ambipolar diffusion.  
We  integrate  Eqs.~(\ref{eq:elect}) 
 up to  $z = 5$ as the intergalactic medium  below 
that redshift is known to be highly ionized from Gunn-Peterson test (see e.g. 
Becker et al. 2001 and references therein).

The modified ionization and thermal history of IGM is quantified in terms
of two observationally measurable parameters: $\tau_{\rm ion}$, $y$. 
\begin{eqnarray}
\tau_{\rm ion} & =  & \int_0^{t_0} n_e \Theta(n_e - n_{es}) \sigma_t c dt \nonumber  \\
y & = & \int_0^{t_0} n_e \sigma_t { k_B (T_e - T_{\rm cbr}) \over m_e c^2}c dt
\label{eq:tauy}
\end{eqnarray}
Here $\tau_{\rm ion}$ measures the optical depth for Thomson scattering in
excess of the standard recombination scenario;
 $n_{es}$ is the number density of electrons for the 
same cosmological model in the absence of magnetic field, $\Theta(n_e - n_{es}) = 1$ if $n_e \ge n_{es}$ and zero otherwise. We assume the 
universe to be fully ionized between the present and $z = 5$, and 
add the $\tau_{\rm ion}$ in this range to the results from magnetic
field decay.  For late 
ionization models, $\tau_{\rm ion}$ reduces to the  
optical depth to the last scattering surface defined in the literature
(see e.g. Bond 1996). Our reason for defining $\tau_{\rm ion}$ differently is that for  the  class of models to be discussed here,  
ionization history begins to differ from the standard case very close 
to the standard recombination epoch. The definition of $\tau_{\rm ion}$ in 
Eq.~(\ref{eq:tauy}) correctly takes that into account. The $y$ parameter 
measures the fractional energy lost/gained by the CMBR which induces spectral 
distortion of CMBR (Zeldovich \& Sunyaev 1969,  Peebles 1993). 
Observations from COBE-FIRAS put stringent upper limit of this parameter: 
$y \la 1.5 \times 10^{-5}$ (Fixsen et al. 1996). 
A non-zero value of $\tau_{\rm ion}$ can be constrained from 
the measurement  of CMBR temperature and polarization anisotropies (see e.g. 
Bond 1996). 

The quantity of direct interest for studying the effect of re-ionization
on the CMBR anisotropies is the visibility function 
$V(z) = (d \tau/d\eta) \exp(-\tau(z))$;
here $d\eta = dt/a$ is the conformal time and 
$\tau = \int_0^\infty n_e \sigma_t c dt $ corresponds to the 
total optical depth not just owing to reionization after the recombination. 
 The visibility functions  measures the normalized probability that a 
photon last scattered between  $\eta$ and $\eta + d\eta$. We show in 
Figure~\ref{fig:f4p} 
visibility functions for the some representative  models in 
comparison with the standard visibility function. We also 
show visibility function for a late re-ionization model in which the 
universe ionizes at $z = 17$ and remains ionized till the present.

Recent  WMAP observation of temperature-polarization cross-correlation
anisotropies suggest that $\tau \simeq 0.17 \pm 0.08$ (Spergel et al. 2003,
Kogut et al. 2003), which means  that between
10 to 20 \% percent photons are re-scattered after the epoch of recombination. 
In view of this observation, we consider  parameter range that
can give $\tau \ga 0.1$ as interesting. 
For the delta function power spectrum
$B_0 \ga 2 \times 10^{-9} \, \rm G$ and $k_\star = k_{\rm max}$
give   $\tau_{\rm ion} \ga 0.1$. The class of power law models we study 
are nearly scale free ($n \simeq -3$) (for detailed discussion see \S~6).
For  these models, $B_0 \ga 3 \times 10^{-9}$ gives $\tau_{\rm ion} \ga 0.1$.
The y-parameter from ambipolar diffusion for the class of models of 
interest  is $\la 2 \times 10^{-7}$, which is much smaller as compared 
to upper limit from  COBE-FIRAS observation.

It should however be borne in mind that CMBR observations are not
only sensitive to the optical depth but also to the shape of visibility 
function (Zaldarriaga 1997). Therefore  our results should not be used directly
to interpret  CMBR observations because, as seen in Figure~\ref{fig:f4p}, 
visibility function in our case gets contribution from a much wider range 
of redshifts than in the case of usual reionization models which have 
been compared with observations. A genuine comparison is possible only
after theoretical predications of  the  CMBR anisotropies 
are computed for our models. A preliminary study shows that our 
models cannot explain the WMAP results, which require a higher 
degree of ionization for $z \la 30$ than our models, 
but generically give new 
anisotropies at smaller scales (larger $\ell$) 
which are distinct from the late reionization models. 
We plan to pursue this issue in more detail in a future study. 

\subsection{The effects of decaying Turbulence}

We now consider the effect of energy input into the IGM due
to decaying MHD turbulence. We calculate this energy input using
Eq.~\ref{eq:dectu} given above.
For our computations we consider a few  cases with $n \simeq -3$ (see \S~6 for
detailed discussion).
 For instance   $n=-2.9$
implies  a decay index $m=0.2/2.1 \sim 0.095$. In all  cases
we adopt $z_i=1000$. We also have to give the ratio
$t_d/t_i$. For this we assume that the decay time-scale
is comparable to the Alfv\'en crossing time associated
with the smallest surviving scales in the magnetic spectrum,
 $k_{\rm max}$. Noting that the time of recombination
$t_i$ corresponds also roughly to the Alfv\'en crossing
time associated with tangles on the magnetic Jeans scale, 
we can take $t_d/t_i \simeq (k_JV_A(k_J,t_i)/(k_{\rm max}V_A(k_{\rm max},t_i))
\simeq (k_J/k_{\rm max})^{(n+5)/2}$. 

Magnetic field dissipation from decaying turbulence is a  faster dissipation
process close to the recombination epoch as compared to the ambipolar 
diffusion. 
This can be the dominant mechanism of magnetic field energy 
dissipation for $k_J < k < k_{\rm max}$. In figures~\ref{fig:f4} 
and~\ref{fig:f5} we show the typical ionization and thermal histories 
for this case. The visibility function for this decay process 
is  shown in Figure~\ref{fig:f4p}. 
The models shown in the figures give $\tau_{\rm ion} \ga 0.1$ and 
a negligibly small $y$ parameter as comparable to observational 
bounds on $y$.  For the magnetic field strengths 
$B_0 \ga 3 \times 10^{-9} \, \rm G$, 
decaying turbulence gives  $\tau_{\rm ion} \ga 0.1$. 

It should be noted that much of the contribution to $\tau_{\rm ion}$ 
comes from close to the epoch of recombination in this case as opposed 
to the magnetic field decay from ambipolar diffusion. And therefore these
two processes will lead to different changes in 
CMBR anisotropies; this can also be seen from from the
plots for the visibility functions shown in Figure~\ref{fig:f4p}. 

In reality both the dissipation mechanisms can act in conjunction.
(As discussed in Section \ref{decay}, for $k > k_J$ modes, 
decaying turbulence is a faster dissipation process close
to recombination, whereas  at redshifts below $z \sim 20$, 
ambipolar diffusion dominates). 
For a field strength $B_0 \simeq  3 \times 10^{-9} \, \rm G$ for nearly
scale invariant model $n = -2.8$, $\tau_{\rm ion} \simeq 0.15$ when
both the dissipation mechanisms are included. As is seen in Figure~3,
 these two dissipation mechanisms affect the 
visibility function at different epochs; their combined effect 
on the visibility function is a function enveloping the curves 
corresponding to both these mechanisms. 

It is of interest to note that both the dissipation mechanisms, and 
especially dissipation due to decaying turbulence, 
lead to changes in the initial 
visibility surface close to the standard recombination.
This can have interesting consequences over and above the changes that 
occur due to late reionization. For instance, the usual late reionization
models lead to decrement in temperature anisotropies for $\ell \ga 40$
by a factor $\exp(-2 \tau)$ and an enhancement of the 
anisotropies for smaller $\ell$ by a similar factor 
(see e.g. Bond 1996). This effect is very
 difficult to discern from
just temperature anisotropies owing to cosmic variance (information about
$\tau$ from WMAP results comes solely from the temperature-polarization
cross-correlation (Kogut et al. 2003)). However a change
in the initial visibility surface can change the diffusion length from
Silk damping (see e.g. Hu \& Sugiyama 1995) which has 
consequences different from the usual late reionization models. 
We hope to study these aspects in detail in the future.

Recent observations of high redshift objects  have shown that the universe
has detectable  amount of neutral hydrogen for  
$z \ge 5.2$ (Becker et al. 2001, Djorgovski et al. 2001, Fan et al. 2002). 
In particular these observations put an upper
bound on the ionized fraction $x_e \la 0.9$ .i.e. at least 10\% universe
becomes neutral in the redshift range  $5.2 \la z \la 6$. This in conjunction
with the the WMAP result suggests that  there might have been two epochs 
of reionization--one occurring as early as $z \simeq 17$ and the other
proceeding for $z \la 5$ (Kogut et al. 2003). The ionization histories 
from magnetic field dissipation (Figures~1 and~5) 
 are compatible with the requirement that more than
10\% of the universe was neutral for $z \ge 5.2$.  

\section{Primordial magnetic fields and early  structure formation}

In the previous two sections we considered the effect of dissipation 
of magnetic fields on the intergalactic medium. Primordial magnetic 
fields also generate density perturbations in the post-recombination 
universe, which can gravitationally collapse to form structures (Wasserman 1978). These structures might  form at high enough redshifts to affect the 
state of intergalactic medium (Kim et al. 1996, SB98a). We consider this 
scenario in this section.

\subsection{Density evolution and power spectrum}

Density evolution in the presence of magnetic field, for scales larger
than the Jeans scale, is governed by equations
(Wasserman 1978, Kim et al. 1996, SB98a, Gopal \& Sethi 2003):
\begin{eqnarray}
{\partial^2\delta_{\scriptscriptstyle b} \over \partial t^2} & = &-2{\dot a \over a} {\partial \delta_{\scriptscriptstyle b} \over \partial t} + 4 \pi G   (\rho_{\scriptscriptstyle \rm DM}\delta_ {\scriptscriptstyle \rm DM} + \rho_{\scriptscriptstyle b} \delta_{\scriptscriptstyle b}) + S(t,{\bf x}) \nonumber \\
{\partial^2\delta_{\scriptscriptstyle \rm DM} \over \partial t^2} & = &-2{\dot a \over a} {\partial \delta_{\scriptscriptstyle \rm DM} \over \partial t} + 4 \pi G   (\rho_{\scriptscriptstyle \rm DM}\delta_ {\scriptscriptstyle \rm DM} + \rho_{\scriptscriptstyle b} \delta_{\scriptscriptstyle b}) 
\label{eq:denevol}
\end{eqnarray}
Here $S(t,{\bf x}) = {\bf \nabla.}({\bf \nabla x B}(t_0,{\bf x})) {\bf x B}(t_0,{\bf x})/(4 \pi \rho_b(t_0) a^3)$ is the source term from magnetic fields; subscript 'DM' refers to dark matter. As seen in Eqs.~(\ref{eq:denevol}),  dark matter perturbations
are not directly affected by the magnetic field, but are generated  by 
baryonic perturbations. 

To solve these equations, 
we define $\delta_{\rm m} = (\rho_{\scriptscriptstyle \rm DM} \delta_{\scriptscriptstyle \rm DM} + \rho_{\scriptscriptstyle \rm b} \delta_{\scriptscriptstyle \rm b})/\rho_{\scriptscriptstyle \rm m}$ 
with $\rho_{\scriptscriptstyle \rm m} = (\rho_{\scriptscriptstyle \rm DM} + \rho_{\scriptscriptstyle \rm b})$. This transforms Eqs.~(\ref{eq:denevol}) to:
\begin{eqnarray}
{\partial^2\delta_{\scriptscriptstyle b} \over \partial t^2} & = &-2{\dot a \over a} {\partial \delta_{\scriptscriptstyle b} \over \partial t} + 4 \pi G   \rho_{\scriptscriptstyle \rm m}\delta_ {\scriptscriptstyle \rm m}  +  S(t,x) \nonumber \\
{\partial^2\delta_{\scriptscriptstyle \rm m} \over \partial t^2} & = &-2{\dot a \over a} {\partial \delta_{\scriptscriptstyle \rm m} \over \partial t} + 4 \pi G   \rho_{\scriptscriptstyle \rm m}\delta_ {\scriptscriptstyle \rm m} +  {\rho_{\scriptscriptstyle b} \over \rho_{\scriptscriptstyle m}}  S(t,x)
\label{eq:denevol1}
\end{eqnarray}
The second of Eqs.~(\ref{eq:denevol1}) 
can be solved by  Green's function method. Its 
solution is:
\begin{equation}
\delta_{\rm \scriptscriptstyle m}(x,t) = A(x) D_1(t) + B(x) D_2(t) - D_1(t)\int_{t_i}^t dt' {S(t',x) D_2(t') \over W(t')} + 
D_2(t)\int_{t_i}^t dt' {S(t',x) D_1(t') \over W(t')}
\label{eq:n5p}
\end{equation}
Here $W(t) = D_1(t)\dot D_2(t) - D_2(t)\dot D_1(t)$ is the Wronskian and 
 $D_1(t)$ and $D_2(t)$ are the solutions of the homogeneous part of the $\delta_{ \rm \scriptscriptstyle m}$ equation (see e.g. Peebles 1980). Here $t_i$ corresponds to the epoch of recombination as the perturbations
cannot grow before this epoch.  The homogeneous
solutions correspond to perturbations generated by sources before 
recombination, e.g. during inflationary epoch. We neglect these
in our analysis. Neglecting homogeneous solutions, the solution to 
Eq.~(\ref{eq:n5p}) for $z \gg 1$ is:
\begin{equation}
\delta_{\rm \scriptscriptstyle m}({\bf x},t) 
\simeq {3 \Omega_b \over 5 \Omega_m^2}\left ({3 \over 2}\left 
({t \over t_i} \right )^{2/3} + \left ({t_i \over t} \right ) - 
{5 \over 2} \right ) S({\bf x},t_i) t_i^2
\label{eq:n5pp}
\end{equation}

Using this solution,  
the evolution of $\delta_{\rm \scriptscriptstyle b}$ can be solved from:
\begin{equation}
{1\over a^2}{\partial \over \partial t} \left (a^2 {\partial \delta_{\rm \scriptscriptstyle b} \over \partial t} \right ) = {3 \over 2} H^2 \delta_{\rm \scriptscriptstyle m} + S(t,{\bf x})
\label{eq:n6p}
\end{equation}
Here we have used $H^2 = (8\pi G /3) \rho_m$. The fastest growing 
solution of Eq.~(\ref{eq:n6p}) for $z \gg 1$  is $\propto  t^{2/3}$. This 
shows that the  
fastest growing modes of both the baryonic  and the dark matter perturbations
grow at the same rate and as can be readily checked from Eqs.~(\ref{eq:n6p})
and~Eqs.~(\ref{eq:n5pp})
are also equal in magnitude.  

The power spectrum of matter  perturbations can written as:
\begin{equation}
P(k,t) = \langle \delta_{\rm \scriptscriptstyle m}(k,t) \delta_{\rm \scriptscriptstyle m}^*(k,t) \rangle \equiv  D^2(t) P(k)
\end{equation}
Here $k$ is the comoving wave vector and
$D(t)$ is dimensionless 
function that gives the time dependence of  the solution to Eq.~(\ref{eq:n6p});
dependences on density etc. are absorbed in the definition of $P(k)$.
For magnetic field power spectrum given in Eq.~(\ref{eq:mag_pow}) and using
Eq.~(\ref{eq:n6}), the matter power spectrum 
 can be evaluated (Kim et al. 1996):
\begin{equation}
P(k) =   {B_0^4 k^3 H_0^{-4}\over (8\pi)^3  \Omega_m^4 \rho_c^2
 k_\star^2} \> \>  \hbox{for} \>  \>  k \le 2 k_\star
\label{eq:matpow}
\end{equation}
For the power law magnetic field  power spectrum, the matter power spectrum
can be computed numerically (Gopal \& Sethi 2003). The matter 
power spectrum cannot grow  below the magnetic Jean's length 
as the magnetic pressure will prevent the structures from collapsing below
this scale (Kim et al. 1996, SB98a).
 This can roughly be taken into account by imposing a sharp
cut-off in the power spectrum below this scale. We adopt the 
more refined analysis of Kim et al. (1996). Kim et~al. (1996)
computed the power spectrum close to the Jeans' length by calculating the 
back-reaction on the velocity field to the first order, this results 
in the modification of the growth factor $D(t)$, which becomes a
function of $k$ for $k \simeq k_J$ and grows more slowly than 
the growing mode of $D(t)$ (Eq.~(\ref{eq:n5pp})) close to the Jeans' scale.
For $k > k_J$, the power spectrum doesn't grow .i.e.  $D(t) = 0$
for these scales and  for scales $k \ll k_J$ the power spectrum grows 
as the usual growing mode (Eq.~(\ref{eq:n5pp})).  Therefore,  the first 
structures to collapse will have scales  
close to the magnetic Jeans' scale $\lambda_J$.
The mass dispersion in a given radius $R$ can then be written as:
\begin{equation}
\sigma^2(R,t) =  4 \pi \int_0^{k_J}  dk k^2 P(k) D^2(k,t)  W^2(kR)
\label{eq:massdis}
\end{equation}
Here $W(kR)$ is the window function. We use the Gaussian
window for our analysis $W(kR) = \exp(-k^2R^2/2)$. Following discussion
preceding Eq.~(\ref{eq:massdis}), we have introduced a cut-off for 
$k > k_J$.  
  As seen in Figures~2 and~6, the 
universe can get heated to temperatures $\simeq 10^4 \, \rm K$, which 
increases the sound speed and therefore 
the Jeans' length of the medium from thermal pressure. At $z \simeq 15$, 
the Jeans' scale from thermal pressure 
corresponds to $k_J^{\rm th} \simeq 200 \, \rm Mpc^{-1}$.
 This corresponds to scales smaller than the magnetic Jeans' length for 
the values of magnetic field we consider here. Therefore, the effect of 
thermal Jeans' length is not important in our analysis.

\subsection{Early structure formation}  

Tangled magnetic fields can 
induce early collapse of structures in the universe. This could also
result in  early re-ionization of the universe.  
The first structures to collapse
would correspond to the length scales
close to magnetic Jeans' length. One can get 
some information about the generic behaviour of 
$\sigma(R)$ from  Eq.~(\ref{eq:massdis}). Firstly  $\sigma \propto R^{-3}$
for $R \ga \lambda_J$ for the delta function magnetic power spectrum,
 which shows that even though first structures 
might form early, the formation of larger structures is suppressed. 
 For  $R = \alpha \lambda_J$ for
 where $\alpha$ is some number typically close   to  one and $k_\star = k_J$ 
 the mass dispersion is nearly independent of 
$B_0$, as  can be established  from Eqs.~(\ref{eq:matpow}) 
and~(\ref{eq:massdis}). (Since for $k_\star =  k_J$, 
$\sigma^2(R,t) \propto B_0^4/(k_{J}^2 \alpha^2 \lambda_J^6) \propto B_0^4 k_J^4/\alpha^2$,
and since $k_J \propto B_0^{-1}$, $\sigma$ becomes independent of $B_0$).

The main dependence of $\sigma(R,t)$ is on the total matter
content of the universe:  $\sigma(R,t) \propto 1/\Omega_m^2$. 
Therefore the redshift at which the first structures collapse
becomes nearly independent  to the value of $B_0$, although the mass
contained in these structures, which depends on the scale 
$k_\star = \alpha k_J$, does depend on $B_0$, through the $k_J$ dependence.

This behaviour is 
generic to most power law magnetic power spectra $M(k) = Ak^n$ 
(Kim et al. 1996, Gopal \& Sethi 2003). Note that the spectral
indices which are of interest from demanding that gravity waves are 
not over produced in the early universe by the anisotropic stress 
of the magnetic fields are $n \simeq -3$, for $B_G \simeq 10^{-9}\, \rm G$ 
(cf. Caprini \& Durrer 2002). 
Further, a spectral index $n$ which is not much larger than $-3$ is
is also required for not over producing
CMB anisotropy on small angular scales (cf. Subramanian and Barrow 2002). 
For $n \le -1.5$, the matter power spectrum is 
$\propto B_0^4 k^{2n+7}/k_c^{2n+6}$ apart from factors proportional 
to $n$ (Gopal \& Sethi 2003), where
$B_0$ now is the rms value filtered at scale $k_c$. This 
gives $\sigma^2(R) \propto B_0^4/ (k_c^{2n+6} R^{2n+10})$. 
For $k_c \simeq k_J$, $\sigma^2(R) \propto B_0^4k_J^4$ again, and 
so does not depend on the value of $B_0$, 
which agrees with the analysis for the delta function 
power spectrum. For $n \simeq -3$, $\sigma(R) \propto 1/R^2$, and again
even though first structures
might form early, the formation of larger structures is suppressed.

In Figure~\ref{fig:f6} and~\ref{fig:f7}  we 
show the  evolution of $\sigma(R,t)$ 
for the delta function magnetic power spectrum and  power 
law spectra with $n \simeq -3$, respectively, for $\Omega_m = 0.3$. 
For a spectrum $P(k)$ with a small scale (large $k$) cut off,
like we have in the present context, 
the scale of the first collapsed structures, $R=R_f$,
would roughly correspond to the wavenumber  
where $k^3P(k)$ is maximum. This maximum occurs at $k \sim k_J$,
but there is always an uncertainty as to whether this corresponds
to $R_f \sim 1/k_J = \lambda_J/(2\pi)$ or $R_f \sim 2\pi/k_J = \lambda_J$.
On the other hand if one uses a Press-Schechter type prescription
to compute the abundance of objects (cf. Padmanabhan 1993), 
then this abundance  will peak for objects where
$d\sigma(R)/dR$ is maximum. 
This criterion gives roughly $R_f \simeq 0.3 - 0.4 \lambda_J$ 
for all the models shown in Figure~\ref{fig:f6} and~\ref{fig:f7} 
(in between the above two estimates based on $k_J$). 
It is in view of this uncertainty in determining the
exact value of $R_f$, that we have given $\sigma(R,t)$ in 
Figure~\ref{fig:f6} and~\ref{fig:f7} for a range around $R=\lambda_J$.

For spherically symmetric perturbations, collapse of a structure 
corresponds to $\sigma(R,t) = 1.68$ (see e.g.  Peebles 1980). Of course
for a power spectrum with a cut-off the first collapses will be more
pancake like. And there is the added complication of
taking account of magnetic pressure effects. Nevertheless it seems reasonable
to demand that $\sigma(R_f,z) \sim 1$ for the formation of
structures. For delta function power spectrum, the collapse of first 
structures then occur for $z \simeq 50$ to $60$. As the  collapse of 
first structures will result in early ionization of the universe, 
this model can probably be ruled out from WMAP observation which 
suggests that the ionization redshift is between
20 and 15. On the other hand for the power law models
with nearly scale invariant spectrum, with say $n \sim -2.8$, 
we see from Figure~\ref{fig:f7}, 
that the collapse redshift of the first 
structures can be in the range suggested by WMAP results. 
Also, from Eq.~(\ref{eq:massdis}), for the 
power law model $\sigma(R) \propto 1/R^2$ and 
therefore only a small fraction of length scales close
to the Jeans' length can collapse early with collapse of 
larger structures suppressed and occurring at a later
redshift (see below). This is in contrast to the 
CDM class of models in which the collapse of first structures leads to collapse
of a much wider range of length scales
(see e.g. Peebles 1993, Padmanabhan 1993). 

Even though the collapse redshift doesn't depend on the value of 
$B_0$, the mass of the collapsed object does depend on $B_0$ through
its dependence on $R_f$. Typical magnetic fields of interest
for CMBR anisotropies are  $B_0$ of order a few nano gauss
(see e.g. Subramanian \& Barrow 2002). 
For $B_0 \sim 3 \times 10^{-9} \, \rm G$ the total mass enclosed within 
a radius $R_f$ is $M_f \simeq 1-3 \times 10^{10} \, \rm M_\odot$, 
which is much smaller than a typical $L_\star$ galaxy. 
For a field $B_0 \sim 10^{-9} \, \rm G$, the mass $M_f$ of the first 
collapsed objects will be smaller, by a factor $ \sim 30$.
Therefore the first structures to collapse 
would be sub-galactic. As we emphasised above, galactic structures,
owing to $1/R^2$ suppression of $\sigma(R)$, form at a much later redshift.
For instance, if the first structures collapse at $z = z_f \simeq 15$, 
the galactic sized objects would only collapse by $z = z_G$ where
$(1+z_G) \sim (R_f/R_G)^2 (1+z_f) = (M_f/M_G)^{2/3}(1+z_f) \ll (1+z_f)$; 
a luminous galaxy of mass $M_G \sim 10^{12} \, \rm M_\odot$ 
would not have collapsed by the present epoch.
This means that even though magnetic fields can induce  the 
formation of first structures, it would have little impact on the 
formation of galactic and larger scale structures at the present 
epoch (see also Gopal \& Sethi 2003). 

From our discussion  above we can conclude that 
(i) collapse of first structures
could have commenced for  $z \simeq 15\hbox{--}30$, (ii) only 
a small fraction of mass range close to the magnetic Jeans' scale collapse 
(iii) the collapse redshift is nearly 
independent of the strength of the magnetic field, if 
the magnetic field is specified as rms filtered at the Jeans scale and
(iv) the mass of the first collapsed objects will be sub-galactic their
exact value depending on $B_0$. These conclusions hold for magnetic 
field strengths for which the magnetic 
Jeans' length exceeds the thermal Jeans' length.  

In light of the recent WMAP 
observations it is of interest to ask if this early collapse can lead to 
early reionization of the universe. This is generally a difficult question 
to address given the uncertainties in  understanding 
astrophysical processes that determine the hydrogen ionizing flux from 
a collapsed structure (see e.g. Kogut et al. 2003 and references therein). 
A quantity of interest for this analysis is the fraction of collapsed 
mass in the universe. In CDM class of models the density field is 
assumed to be Gaussian, which allows the collapsed fraction to be computed
from Press-Schechter method (see e.g. Padmanabhan 1993). The collapsed 
fraction of mass at scales which go non-linear generally approaches a 
high fraction of unity in the Gaussian case. In the magnetic field induced 
structure formation, it is the magnetic field that is assumed to have 
Gaussian statistics. The density field is $\propto B^2$ and satisfies
modified $\chi^2$ statistics (cf. Subramanian and Barrow 1998). 
In this case the high density regions get 
more weight than a Gaussian, and therefore it is natural to expect that 
the formation  of  first 
structures would result in the collapse of a 
large fraction of the mass in the  universe. 
It is then possible that the magnetically induced collapse of early 
structures have interesting consequences for the reionization of the 
universe.  We hope to return to this issue in a later work. 

\section{Conclusions}

We have studied here some of the consequences 
of primordial tangled magnetic fields for the post-recombination
universe. In the post-recombination epoch the  magnetic field energy
can dissipate into the intergalactic medium by  ambipolar diffusion
and by generating decaying MHD turbulence. 
An important issue we focussed upon, is 
the impact of such dissipation on the 
thermal and ionization history of the universe.
We showed that magnetic field dissipation can change the 
ionization history of the universe sufficiently to have a bearing on 
recent and future observations of CMBR anisotropy. 
We have also shown that primordial magnetic fields
generically induce the early formation of sub-galactic structures, which 
could be responsible for the reionization of the universe at $z \simeq 15$, as 
indicated by recent WMAP results (Kogut et al. 2003).

More specifically, our results are as follows:
\begin{itemize}
\item[1.] Primordial magnetic field dissipation  can 
result in ionization histories that give 
the Thompson scattering optical depth $\tau_{\rm ion} \ga 0.1$. 
This requires magnetic field strengths 
$B_0 \ga 2 \times 10^{-9} \, \rm G$, with $k_\star = k_{max}$,
for the delta function power spectrum. 
For nearly scale invariant power law power spectra 
magnetic field strengths (smoothed over $k_{max}$), 
$B_0 \ga 3 \times 10^{-9} \, \rm G$ gives $\tau_{\rm ion} \ga 0.1$   
from either of  the dissipation processes 
(ambipolar diffusion and decaying MHD 
turbulence). Adding the effect of both dissipation mechanisms
give  $\tau_{\rm ion} \ga 0.15$, for the same field strength.
(Note that these two effects can indeed add since 
they are important at different epochs; the major
effect of decaying turbulence occurring at high $z$ after
recombination, while ambipolar diffusion dominates at
redshifts below $z \simeq  20$.)

\item[2.] To infer the impact of these dissipation processes,
on CMBR measurements, we have computed the visibility 
function for the resulting ionization histories. 
Our preliminary analysis show that
the  recent WMAP observations are unlikely to have been much affected
by the magnetic field dissipation, even though the Thomson scattering 
optical depth is comparable to the value  inferred by WMAP. This is because 
the visibility function in the case of magnetic field dissipation receives
contribution from a much wider range of redshifts than is required
to explain the WMAP observations. Future CMBR probes like Planck
can potentially detect the modified CMBR anisotropy signal from 
such partial re-ionization (Kaplinghat et al. 2003). 
This can be used to detect or further constrain small scale 
primordial magnetic fields.

\item[3.] Primordial magnetic fields can induce the collapse of 
first subgalactic structures in the universe at high redshifts.
We show that for the nearly scale free power law models the collapse 
redshift is in the range between $10$ to $20$.
The masses of these objects depend on the magnetic field strength
smoothed on the Jeans scale, and lie in the range 
$3 \times 10^9 M_\odot$ to $3 \times 10^{10} M_\odot$,
for $B_0 \sim 10^{-9} \, \rm G$ to $B_0 \ga 3 \times 10^{-9} \, \rm G$.
As the formation of first structures is likely to ionize the 
universe, it seems plausible that the formation of these objects
could explain the reionization features
implied by WMAP results (Kogut et al. 2003) .  
\end{itemize}

In summary, the presence of tangled magnetic fields in the post-recombination
epoch can result in several interesting signals.
They can induce early collapse of structures, which might explain 
the recent WMAP polarization results. 
In addition, the slow dissipation of magnetic field
in the post-recombination era can alter the ionization history sufficiently
to have interesting observational consequences  especially for  the 
future CMBR probes like Planck.  

\section*{Acknowledgment}
We  would like to thank S. Sridhar for many fruitful
discussions.

\newpage

\begin{figure}
\epsfig{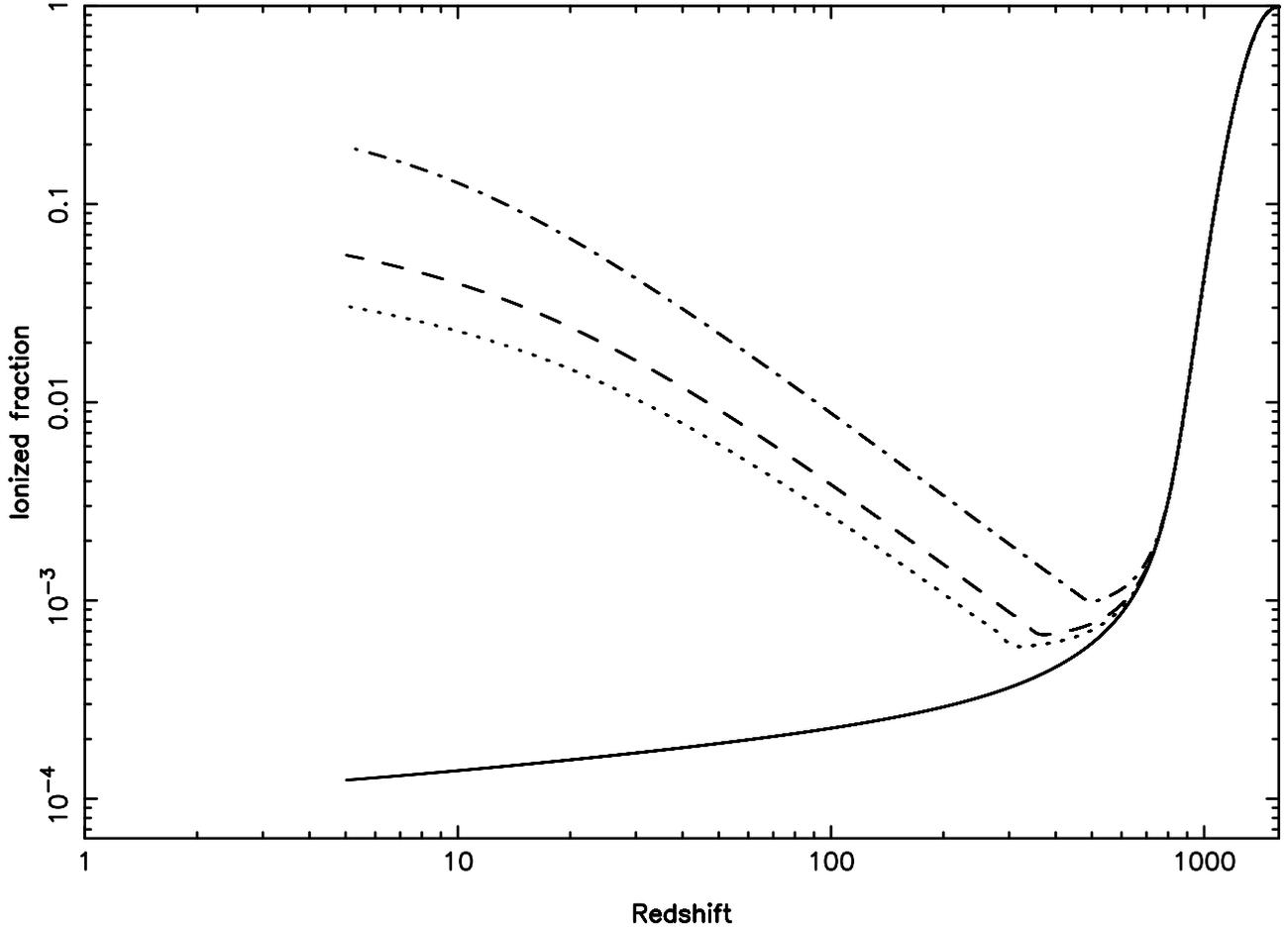}
\caption{Evolution of the ionization state of the universe is
shown for ambipolar dissipation. Different curves are:
standard recombination (solid curve); the dotted and  dashed curves
correspond to nearly scale free magnetic field power spectra with 
$n = -2.9$ and $n = -2.8$ with $B_0 = 3 \times 10^{-9} \, \rm G$;   the
dot-dashed curves correspond to the delta function magnetic field power spectrum
with  $B_0 = 3 \times 10^{-9} \, \rm G$ and $k_\star = k_{\rm max}$. 
}
\label{fig:f1}
\end{figure}

\begin{figure}
\epsfig{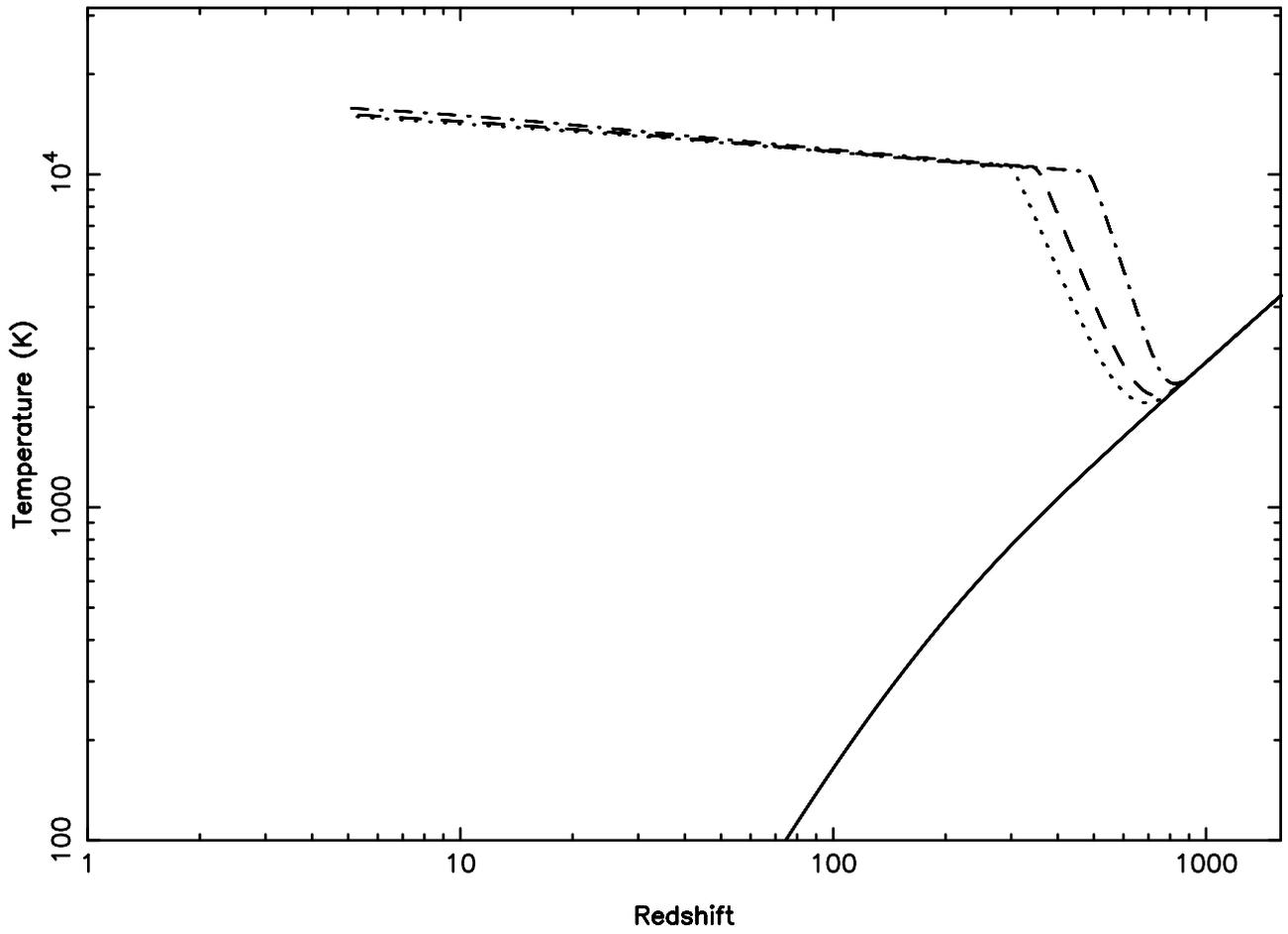}
\caption{Evolution of the thermal  state of the universe is
shown for ambipolar dissipation. Curves are for the same 
 parameters as in 
Figure~1
}
\label{fig:f2}
\end{figure}

\begin{figure}
\epsfig{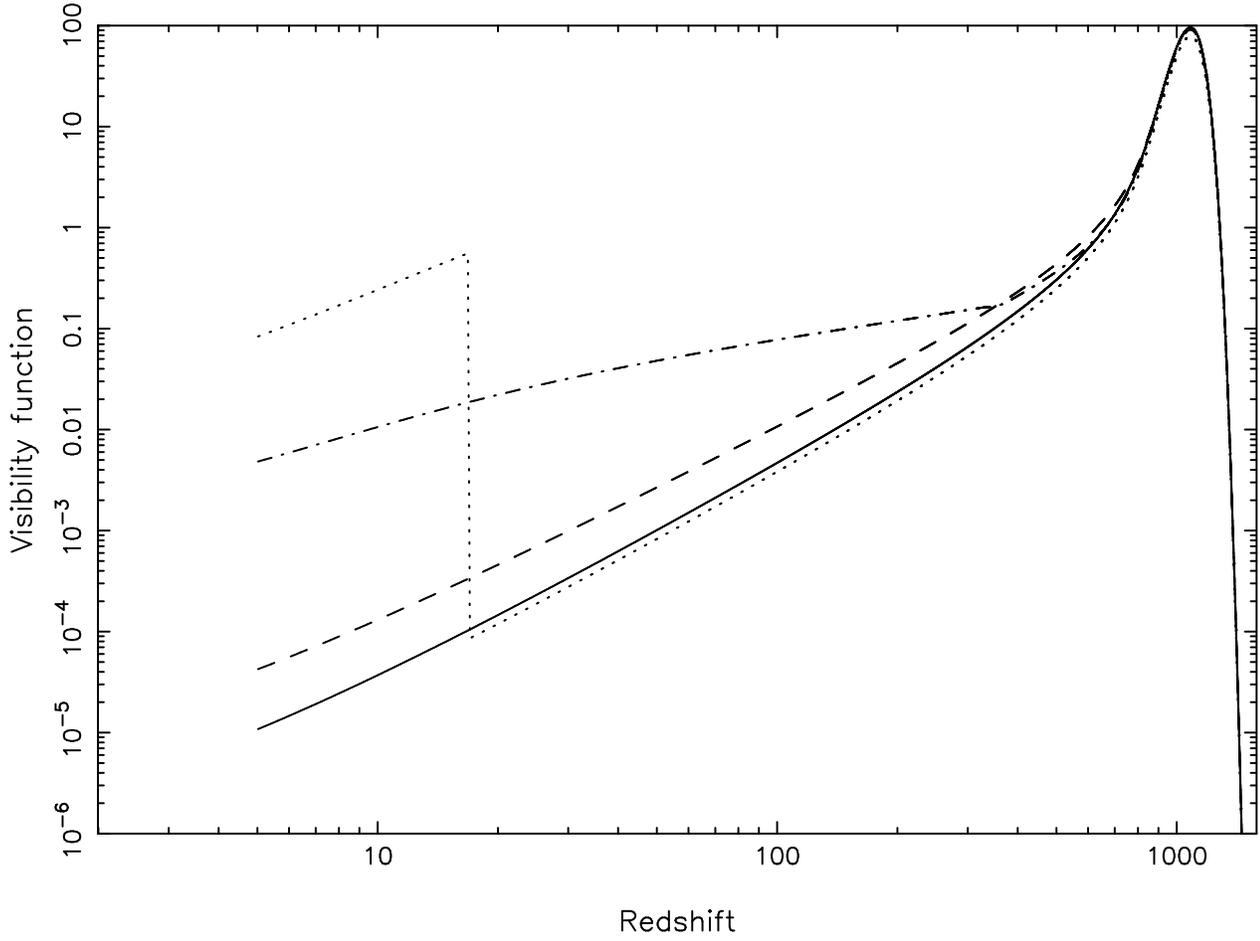}
\caption{Visibility function, defined as
$d\tau/d\eta \exp(-\tau) H_0^{-1}$, is plotted for different models. The 
solid and the dotted  curves are for the standard recombination 
and a model in which
the universe reionizes at $z =17$, respectively.
 Dashed curve
corresponds  to a  decaying turbulence model with   $B_0 = 3 \times 10^{-9} \, \rm G$ and $m = 0.2$ (Eq.~({\ref{eq:dectu})). Dot-dashed  curve
corresponds to the 
ambipolar diffusion case with $B_0 = 3 \times 10^{-9} \, \rm G$
and $n = -2.8$.}}
\label{fig:f4p}
\end{figure}

\begin{figure}
\epsfig{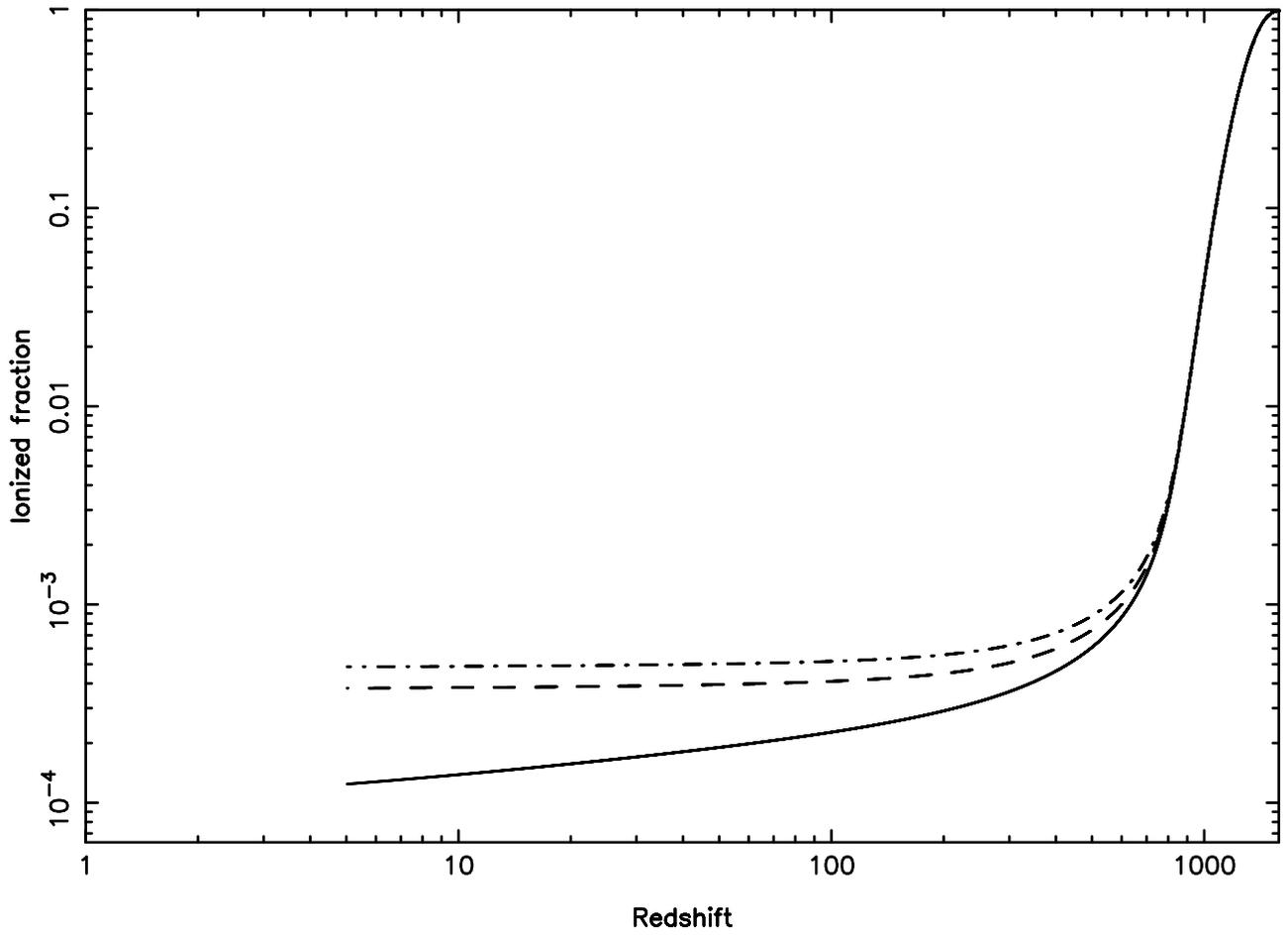}
\caption{Evolution of the ionization state of the universe is
shown for  dissipation of magnetic field energy due to
decaying turbulence. Different curves are:
standard recombination (solid curve), the  dot-dashed, and 
dashed curves correspond to 
$B_0 = 3 \times 10^{-9} \, \rm G$, $m = \{0.2,0.1\}$, respectively (Eq.~({\ref{eq:dectu})). 
}}
\label{fig:f4}
\end{figure}

\begin{figure}
\epsfig{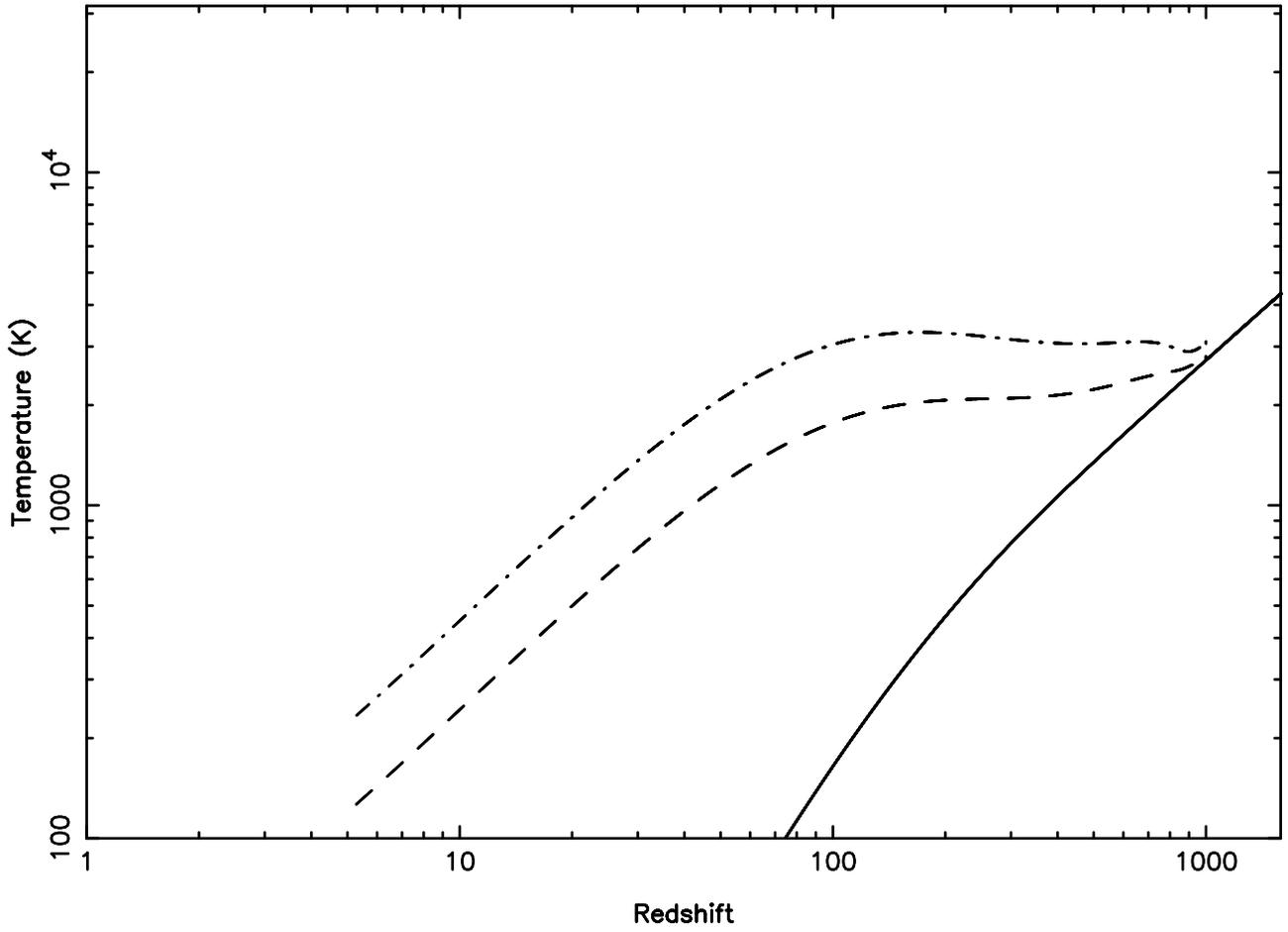}
\caption{Evolution of the thermal  state of the universe is
shown for decaying turbulence. Curves are for the same 
 parameters as in 
Figure~\ref{fig:f4}
}
\label{fig:f5}
\end{figure}

\begin{figure}
\epsfig{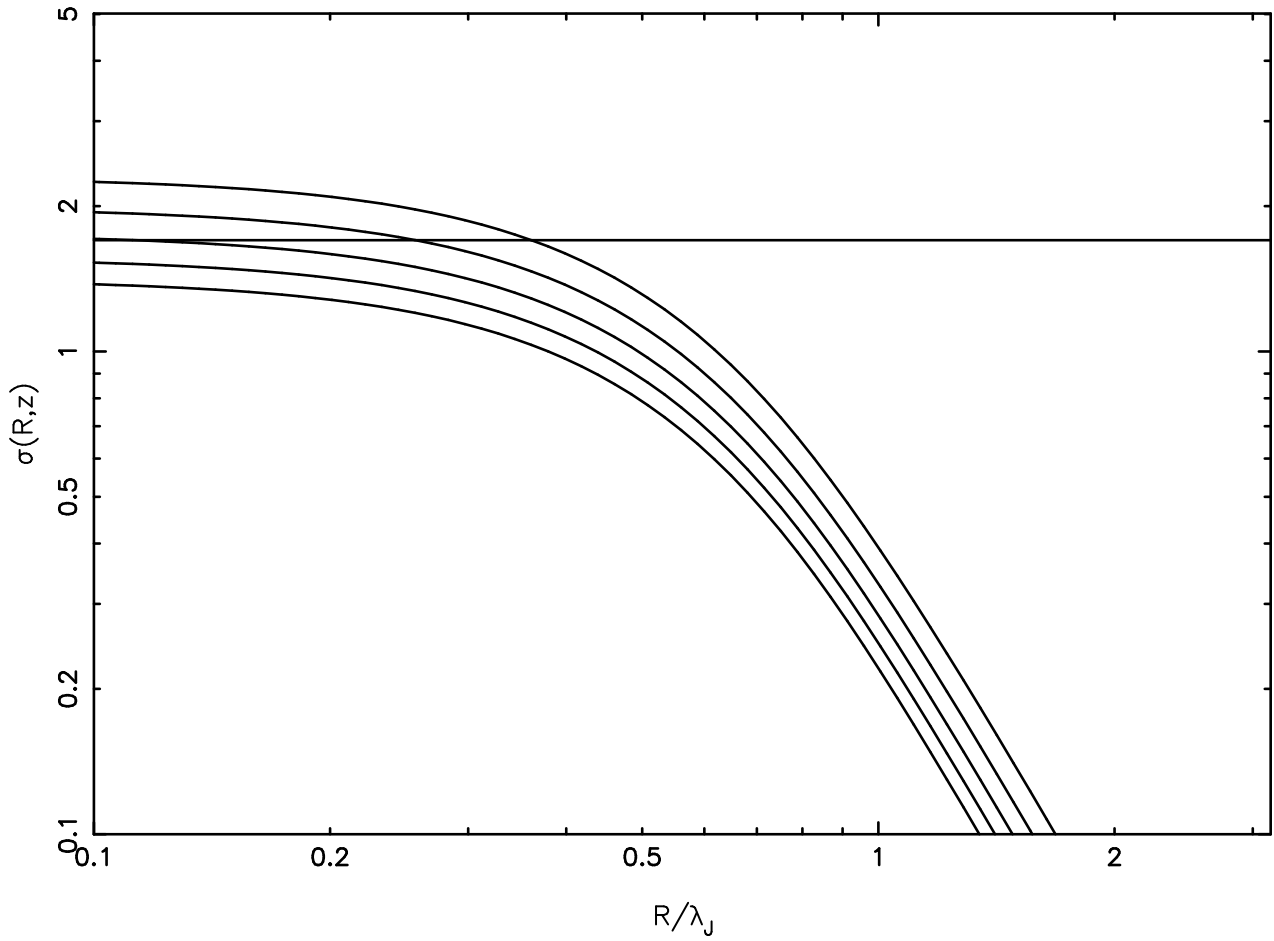}
\caption{The mass dispersion $\sigma(R,z)$ is shown for the delta
function magnetic field power spectrum. Different curves, from top to
bottom,  correspond to redshifts $z = \{50,60,70,80,90\}$, respectively.
 The horizontal line corresponds to $\sigma = 1.68$. 
}
\label{fig:f6}
\end{figure}

\begin{figure}
\epsfig{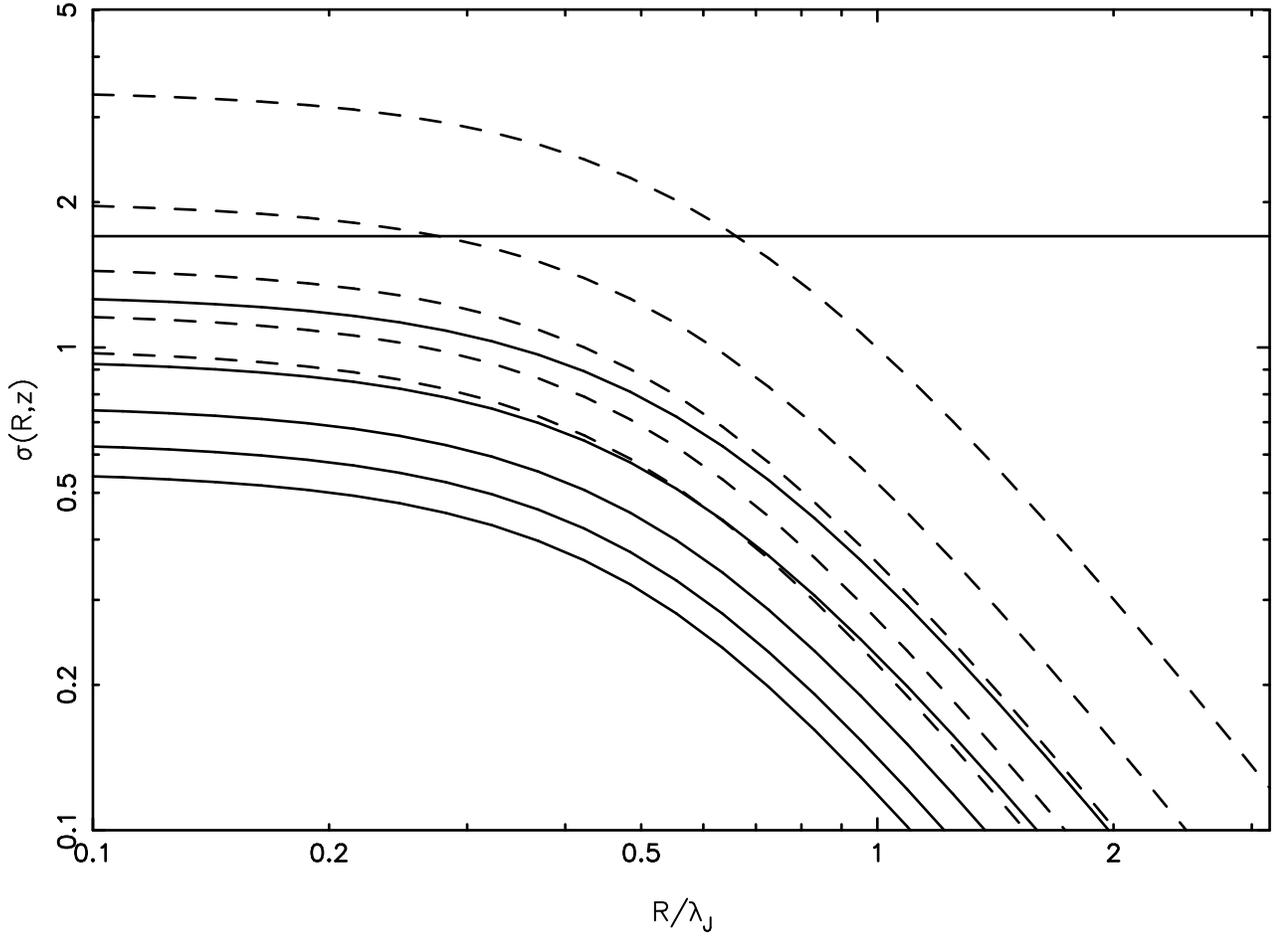}
\caption{The mass dispersion $\sigma(R,z)$ is shown for two
models with nearly scale free  magnetic field power spectra. The solid 
and dashed curves correspond to $n = -2.9$ and $n = -2.8$, respectively.
   Different curves, from top to
bottom,  correspond to redshifts $z = \{10,15,20,25,30\}$, respectively.
 The horizontal line corresponds to $\sigma = 1.68$. 
}
\label{fig:f7}
\end{figure}


\begin{thebibliography}{}
\bibitem[]{} Banerjee, R. \& Jedamzik, K, 2003, Physical Review Letters, 91,
251301-1
\bibitem[]{} Banerjee, R. 2003, talk at COSMO 2003
\bibitem[Barrow, Ferreira, \& Silk(1997)]{1997PhRvL..78.3610B} Barrow, 
J.~D., Ferreira, P.~G., \& Silk, J.\ 1997, Physical Review Letters, 78, 
3610 
\bibitem[]{}  Beck R., Brandenburg A., Moss D., Shukurov A. M., Sokoloff
D. D., 1996, Ann. Rev. Astron. Astrophys., 34, 155
\bibitem[Becker et al.(2001)]{2001AJ....122.2850B} Becker, R.~H.~et al.\ 
2001, AJ, 122, 2850 
\bibitem[]{BB02} Blackman, E. G. \& Brandenburg, A., 2002, 579, 359
\bibitem[]{BF00}  Blackman E. G. \&  Field G. F., 2000, ApJ, 534, 984
\bibitem[]{B01}  Brandenburg A., 2001, ApJ, 550, 824
\bibitem[]{BS00}  Brandenburg A., Subramanian K., 2000, A\&A, 361, L33
\bibitem[]{} Brandenburg, A. \& Subramanian, K., 2004, astro-ph/0405052 
\bibitem[Schaeffer, Silk, Spiro, \& Zinn-Justin(1996)]{1996clss.conf.....S}Bond, R., in  Schaeffer, R., Silk, J., Spiro, M., \& Zinn-Justin, J.\ 1996, ASP
\bibitem[Brandenburg, Enqvist, \& Olesen(1996)]{1996PhRvD..54.1291B} 
Brandenburg, A., Enqvist, K., \& Olesen, P.\ 1996, Phys. Rev. D, 54, 1291
\bibitem[Caprini \& Durrer(2002)]{2002PhRvD..65b3517C} Caprini, C.~\& Durrer, R.\ 2002, \prd, 65, 23517 
\bibitem[Carilli \& Taylor(2002)]{2002ARA&A..40..319C} Carilli, C.~L.~\& Taylor, G.~B.\ 2002, \araa, 40, 319 
\bibitem[]{cat}  Cattaneo F., Vainshtein S. I., 1991, ApJ, 376, L21
 \bibitem[Christensson, Hindmarsh, \& Brandenburg(2001)]{2001PhRvE..64e6405C} Christensson, M., Hindmarsh, M., \& Brandenburg, A.\ 2001, Phys. Rev. E,
 64, 56405 
\bibitem[Clarke, Kronberg, \& B{\" o}hringer(2001)]{2001ApJ...547L.111C} 
Clarke, T.~E., Kronberg, P.~P., \& B{\" o}hringer, H.\ 2001, \apjl, 547, 
L111 
\bibitem[]{} Cowling, T. G. 1956, MNRAS, 116, 114
\bibitem[Djorgovski, Castro, Stern, \& Mahabal(2001)]{2001ApJ...560L...5D} 
Djorgovski, S.~G., Castro, S., Stern, D., \& Mahabal, A.~A.\ 2001, \apjl, 
560, L5 
\bibitem[]{dur}  Durrer R., Ferreira P. G., Kahniashvili T., 2000, \prd,
61, 043001
\bibitem[Fan et al.(2002)]{2002AJ....123.1247F} Fan, X., Narayanan, V.~K., 
Strauss, M.~A., White, R.~L., Becker, R.~H., Pentericci, L., \& Rix, H.\ 
2002, AJ, 123, 1247 
\bibitem[Fixsen et al.(1996)]{1996ApJ...473..576F} Fixsen, D.~J., Cheng, 
E.~S., Gales, J.~M., Mather, J.~C., Shafer, R.~A., \& Wright, E.~L.\ 1996, 
\apj, 473, 576 
\bibitem[Freedman et al.(2001)]{2001ApJ...553...47F} Freedman, W.~L.~et 
al.\ 2001, \apj, 553, 47
\bibitem[]{MG03}  Giovannini, M, 2004, Int.J.Mod.Phys., D13, 391 
\bibitem[]{} Gopal, R. \& Sethi, S. K., 2003, Journal of Astrophysics 
and Astronomy, 24, 51
\bibitem[]{rubgras}  Grasso D., Rubinstein H. R., 2001, Phys. Rep., 348, 161
\bibitem[]{GD}  Gruzinov A. V., Diamond P. H., 1994, PRL, 72, 1651
\bibitem[Harrison(1970)]{1970MNRAS.147..279H} Harrison, E.~R.\ 1970, 
\mnras, 147, 279 
\bibitem[Hu \& Sugiyama(1995)]{1995ApJ...444..489H} Hu, W.~\& Sugiyama, N.\ 
1995, \apj, 444, 489 
\bibitem[Jedamzik, Katalini{\' c}, \& Olinto(1998)]{1998PhRvD..57.3264J} 
Jedamzik, K., Katalini{\' c}, V., \& Olinto, A.~V.\ 1998, \prd, 57, 3264 
\bibitem[Kaplinghat et al.(2003)]{2003ApJ...583...24K} Kaplinghat, M., Chu, 
M., Haiman, Z., Holder, G.~P., Knox, L., \& Skordis, C.\ 2003, \apj, 583, 
24   
\bibitem[Kim, Olinto, \& Rosner(1996)]{1996ApJ...468...28K} Kim, E., 
Olinto, A.~V., \& Rosner, R.\ 1996, \apj, 468, 28
\bibitem[Kim, Kronberg, Giovannini, \& Venturi(1989)]{1989Natur.341..720K} 
Kim, K.-T., Kronberg, P.~P., Giovannini, G., \& Venturi, T.\ 1989, \nat, 
341, 720 
\bibitem[]{klee}  Kleeorin N., Moss D., Rogachevskii I., Sokoloff D., 2000,
A\&A, 361, L5
\bibitem[]{}  Kogut, A.  et al, 2003, ApJS, 148, 161
\bibitem[]{ka}  Kulsrud R. M., Anderson S. W., 1992, ApJ, 396, 606
\bibitem[Kulsrud, Cen, Ostriker, \& Ryu(1997)]{1997ApJ...480..481K} 
Kulsrud, R.~M., Cen, R., Ostriker, J.~P., \& Ryu, D.\ 1997, \apj, 480, 481 
\bibitem[]{} Landau, L. D.  \& Lifshitz, E. M. 1987, Fluid Mechanics, Pergamon Press
\bibitem[]{} Madau, P., Meiksin, A., \& Rees, M. J. 1997, 475, 429
\bibitem[Mestel \& Spitzer(1956)]{1956MNRAS.116..503M} Mestel, L.~\& Spitzer, L.\ 1956, \mnras, 116, 503 
\bibitem[M{\" u}ller \& Biskamp(2000)]{2000PhRvL..84..475M} M{\" u}ller, 
W.~\& Biskamp, D.\ 2000, Physical Review Letters, 84, 475 
\bibitem[]{} Olesen, P., 1997, Phys. Lett., B398, 321
\bibitem[Oren \& Wolfe(1995)]{1995ApJ...445..624O} Oren, A.~L.~\& Wolfe, 
A.~M.\ 1995, \apj, 445, 624 
\bibitem[Padmanabhan(1993)]{1993sfu..book.....P} Padmanabhan, T.\ 1993, 
Cambridge, UK: Cambridge University Press
\bibitem[Parker(1979)]{1979cmft.book.....P} Parker, E.~N.\ 1979, Oxford, 
Clarendon Press; New York, Oxford University Press, 1979, 858 p.
\bibitem[]{} Peebles, P.~J.~E.\ 1993, 
Principles of Physical Cosmology, Princeton University Press
\bibitem[]{} Peebles, P.~J.~E.\ 1980, The Large Scale Structure of the Universe, Princeton University Press
\bibitem[]{} Peebles, P.~J.~E.\ 1968, \apj, 153, 1
\bibitem[Perlmutter et al.(1999)]{1999ApJ...517..565P} Perlmutter, S.~et 
al.\ 1999, \apj, 517, 565
\bibitem[Ratra(1992)]{1992ApJ...391L...1R} Ratra, B.\ 1992, \apjl, 391, L1
\bibitem[Rees \& Reinhardt(1972)]{1972A&A....19..189R} Rees, M.~J.~\& Reinhardt, M.\ 1972, \aap, 19, 189 
\bibitem[]{} Ricotti, M. \& Ostriker, J. P. 2003, astro-ph/0311003
\bibitem[]{} Riess, A.~G.~et al.\ 2004  astro-ph/0402512
\bibitem[Riess et al.(1998)]{1998AJ....116.1009R} Riess, A.~G.~et al.\ 
1998, \aj, 116, 1009
\bibitem[]{}  Ruzmaikin A. A., Shukurov A. M., Sokoloff D. D., 1988, {\it %
Magnetic Fields of Galaxies}, Kluwer, Dordrecht (1988)
\bibitem[Seshadri \& Subramanian(2001)]{2001PhRvL..87j1301S} Seshadri, 
T.~R.~\& Subramanian, K.\ 2001, Physical Review Letters, 87, 101301 
\bibitem[]{} Sethi, S. K., 2003, \mnras, 342, 962
\bibitem[]{} Shu, F. H. 1992, Gas Dynamics, University Science Books
\bibitem[]{} Shiromizu, T., 1998, Phys. Lett., B443, 127
\bibitem[]{shukurov04} Shukurov, A., 2004, Introduction to galactic dynamos,
In Mathematical aspects of natural dynamos, Ed. E. Dormy,
Kluwer Acad. Publ., Dordrecht.
\bibitem[]{} Spergel, D. N. et al. 2003, ApJS, 148, 175
\bibitem[]{suba}  Subramanian K., 1998, MNRAS, 294, 718
\bibitem[]{ks99}  Subramanian K., 1999, PRL, 83, 2957
\bibitem[]{} Subramanian, K., 2002, Bull.\ Astr.\ Soc.\ India, 30, 715
\bibitem[Subramanian \& Barrow(1998)]{1998PhRvL..81.3575S} Subramanian, 
K.~\& Barrow, J.~D.\ 1998, Physical Review Letters, 81, 3575 
\bibitem[Subramanian \& Barrow(1998)]{1998PhRvD..58h3502S} Subramanian, 
K.~\& Barrow, J.~D.\ 1998, \prd, 58, 83502 (SB98a)
\bibitem[Subramanian \& Barrow(2002)]{2002MNRAS.335L..57S} Subramanian, 
K.~\& Barrow, J.~D.\ 2002, \mnras, 335, L57 
\bibitem[Subramanian, Narasimha, \& Chitre(1994)]{1994MNRAS.271L..15S} 
Subramanian, K., Narasimha, D., \& Chitre, S.~M.\ 1994, \mnras, 271, L15 
\bibitem[Subramanian, Seshadri, \& Barrow(2003)]{2003MNRAS.344L..31S} 
Subramanian, K., Seshadri, T.~R., \& Barrow, J.~D.\ 2003, \mnras, 344, L31 
\bibitem[]{} Tonry, J. L. et al.~2003, \apj, 594, 1
\bibitem[Turner \& Widrow(1988)]{1988PhRvD..37.2743T} Turner, M.~S.~\& Widrow, L.~M.\ 1988, \prd, 37, 2743
\bibitem[Tytler, O'Meara, Suzuki, \& Lubin(2000)]{2000PhR...333..409T} 
Tytler, D., O'Meara, J.~M., Suzuki, N., \& Lubin, D.\ 2000, \physrep, 333, 
409 
\bibitem[]{vogt_ensslin03} Vogt, C. \& Ensslin, T. A., 2003, \aap, 412, 373
\bibitem[Wasserman(1978)]{1978ApJ...224..337W} Wasserman, I.\ 1978, \apj, 
224, 337
\bibitem[Widrow(2002)]{2002RvMP...74..775W} Widrow, L.~M.\ 2002, Reviews of
Modern Physics, vol.~74, Issue 3, pp.~775-823, 74, 775
\bibitem[Zaldarriaga(1997)]{1997PhRvD..55.1822Z} Zaldarriaga, M.\ 1997, 
\prd, 55, 1822
\bibitem[Zeldovich, Ruzmaikin, \& Sokolov(1983)]{1983flma....3.....Z} 
Zeldovich, I.~B., Ruzmaikin, A.~A., \& Sokolov, D.~D.\ 1983, New York, 
Gordon and Breach Science Publishers (The Fluid Mechanics of Astrophysics 
and Geophysics.~Volume 3), 1983, 381 p.
 \bibitem[Zeldovich \& Sunyaev(1969)]{1969Ap&SS...4..301Z} Zeldovich, 
Y.~B.~\& Sunyaev, R.~A.\ 1969, Ap \& SS, 4, 301 
\bibitem[]{} Zel'dovich, Ya. B., Kurt, V. G., \& Sunyaev, R. A. 1969, Soviet Phys. JETP, 28, 146

\end{thebibliography}
\end{document}